%% file: krause_et_al_2012a.tex
\documentclass[structabstract]{aa}  
\usepackage{mathptmx}
\usepackage{amssymb}
\usepackage{natbib}
\usepackage{graphicx}
\usepackage{color}
\usepackage{txfonts}
\input{def}

\newcommand{\rem}[1]{}
\newcommand{\ch}[1]{#1}
\newcommand{\revtwo}[1]{{ #1}}
\newcommand{\rev}[1]{#1}
\begin{document}
\title{Feedback by massive stars and the emergence of superbubbles}
\titlerunning{Emergence of superbubbles -- energy efficiency \&
  Vishniac instabilities}
\subtitle{I. Energy efficiency \& Vishniac instabilities}

\author{Martin Krause \inst{1,2} \fnmsep\thanks{E-mail:
    Martin.Krause@universe-cluster.de} \and \ch{Katharina Fierlinger}
  \inst{3,1} \and Roland Diehl \inst{2,1} \and Andreas Burkert
  \inst{3,2,4} \and Rasmus Voss \inst{5} \and
  Udo Ziegler \inst{6} }

\institute{Excellence Cluster Universe, Technische Universit\"at
  M\"unchen, Boltzmannstrasse 2, 85748 Garching, Germany \and
  Max-Planck-Institut f\"ur extraterrestrische Physik, Postfach 1312,
  Giessenbachstr., 85741 Garching, Germany \and
  Universit\"atssternwarte M\"unchen, Scheinerstr.~1, 81679 M\"unchen,
  Germany \and Max-Planck-Fellow \and Department of
  Astrophysics/IMAPP, Radboud University Nijmegen, PO Box 9010,
  NL-6500 GL Nijmegen, the Netherlands \and Leibniz-Institut f\"ur
  Astrophysik Potsdam (AIP), An der Sternwarte 16, D-14482 Potsdam }

\date{Submitted: July 20, 2012; accepted: December 10, 2012 }

 
\abstract
{Massive stars influence their environment via stellar winds, ionising
  radiation and supernova explosions. This is signified by observed
  interstellar bubbles. Such ``feedback'' is an important factor for
  galaxy evolution theory and galactic wind models. The efficiency of
  the energy injection into the interstellar medium via bubbles and
  superbubbles is uncertain, and is usually treated as a free
  parameter for galaxy scale effects. 
\ch{In particular, since many stars are born in groups it is
  interesting to study the dependence of the effective energy
  injection on the concentration of the stars.}}
{We aim to reproduce observations of superbubbles, their relation to
  the energy injection of the parent stars and to understand their
  effective energy input into the interstellar medium (ISM)\ch{, as a
    function of the spatial configuration of the group of parent stars}. }
{We study the evolution of isolated and merging interstellar bubbles
\ch{of three stars (25, 32 and 60~$M_\odot$) in a homogeneous
  background medium with a density of 10~$m_\mathrm{p}$~cm$^{-3}$ }
  via 3D-hydrodynamic simulations with standard ISM thermodynamics
  (optically thin radiative cooling and photo-electric heating) and
  time dependent energy and mass input according to stellar
  evolutionary tracks. \ch{We vary the position of the three stars
    relative to each other to compare the energy response for cases of
    isolated, merging and initially cospatial bubbles.}}
{Due to mainly the Vishniac instability, our simulated bubbles develop thick shells and filamentary internal
  structures in column density. The
  shell widths reach tens of per cent of the outer bubble radius, 
  which compares favourably to
  observations.  More energy is retained in the ISM for more closely
  packed groups\ch{, by up to a factor of three and typically a factor
  of two for intermediate times after the first supernova}. 
  Once the superbubble is established, different
  positions of the contained stars make only \ch{a} minor difference to the
  energy tracks.  \ch{For our case of three massive stars, the energy
  deposition varies only very little for distances up to about 30~pc
  between the stars.}
  Energy injected by supernovae is entirely dissipated
  in a superbubble on a timescale of about 1~Myr\ch{, which increases
    slightly with the superbubble size at the time of the
    explosion.} }
{\ch{The Vishniac instability may be responsible for the broadening
  of the shells of interstellar bubbles. Massive star winds are
  significant energetically due to their -- in the long run -- more
  efficient, steady energy injection and because they evacuate the
  space around the massive stars. 
  For larger scale simulations, the feedback effect of
  close groups of stars or clusters may be subsumed  into one
  effective energy input with insignificant loss of energy accuracy.}
}


\maketitle
%

\section{Introduction}\label{intro}

\begin{figure}
  \centering
  \includegraphics[width=0.5\textwidth]{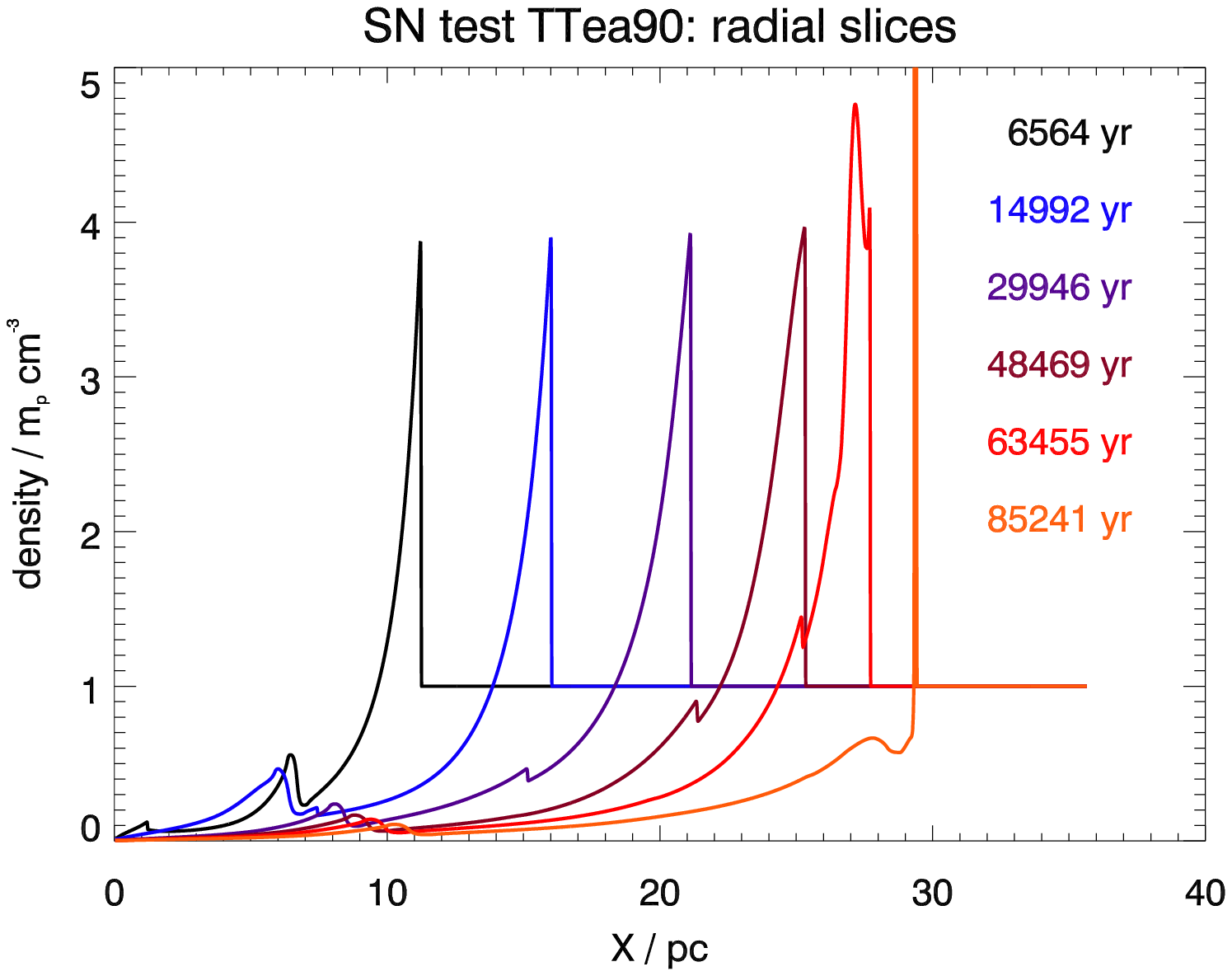}
  \includegraphics[width=0.5\textwidth]{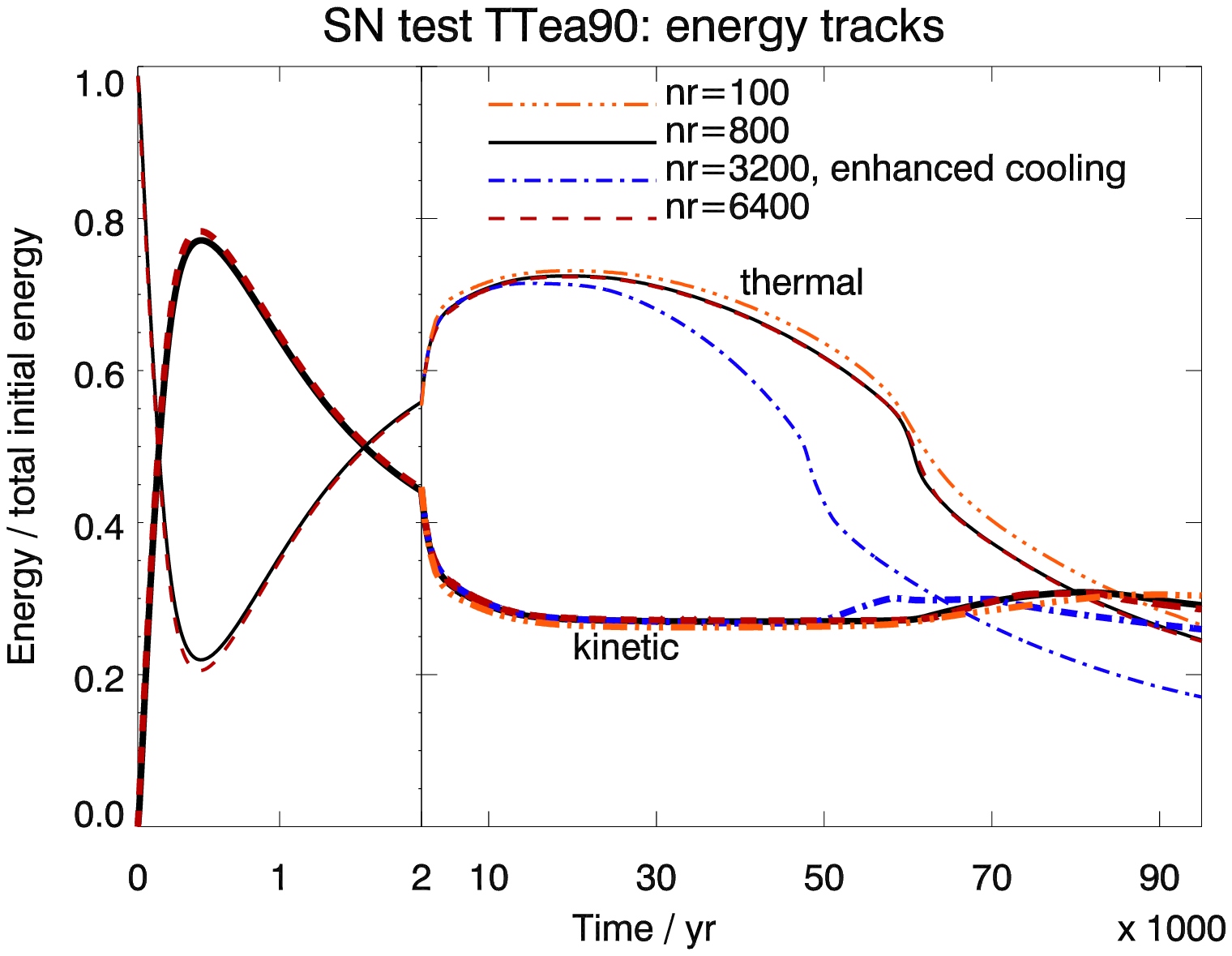}
  \includegraphics[width=0.5\textwidth]{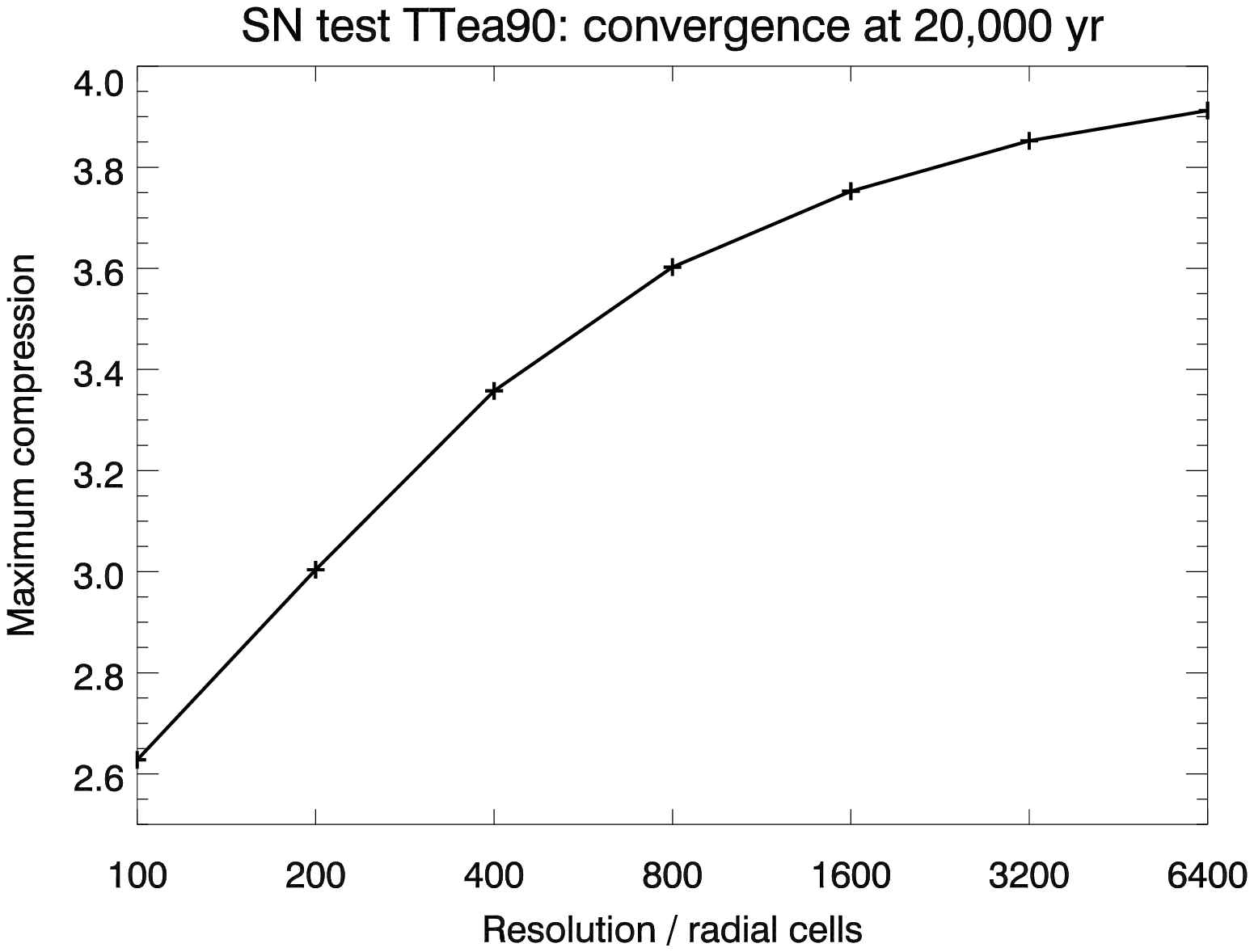}
  \caption{Test simulation of an isolated supernova in a homogeneous
    environment according to \citet{Tenea90}. 
    \revtwo{This test was run in 1D and spherical coordinates at
      radial resolutions between 100 and 6400 cells with respective
      increments of a factor of two and with the standard cooling and
      heating. Top:
    density slices at different times (increasing from left 
    to right). Middle: evolution of thermal (thin lines)
    and kinetic (thick lines) energy for the 100 (orange triple
    dot-dashed lines), 800 (solid black lines), 6400 (red dashed lines)
    radial cells simulation. Additionally a simulation with 3200
    radial cells and enhanced cooling is shown (blue  dot-dashed
    lines, see text for details). 
    The left part of the plot
    zooms into the first 2000 yr of the evolution. Bottom: Maximum
    compression at about 20,000~yr as a function of resolution. The
    expected maximum compression at this time is four for a strong
    adiabatic shock.}}
  \label{f:TTtest}%
\end{figure}
%
%
\begin{figure}
  \centering
  \includegraphics[width=0.5\textwidth]{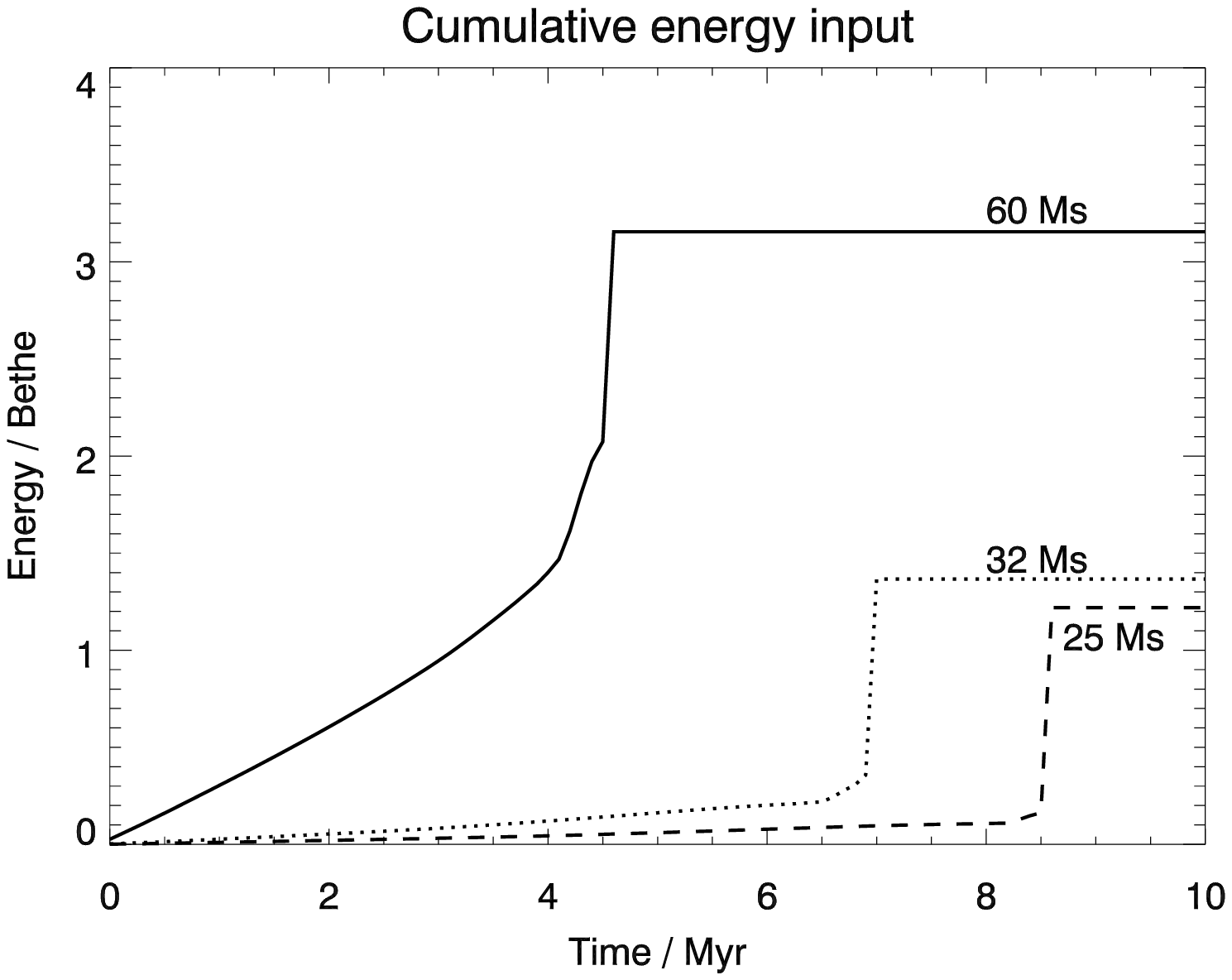}
  \includegraphics[width=0.5\textwidth]{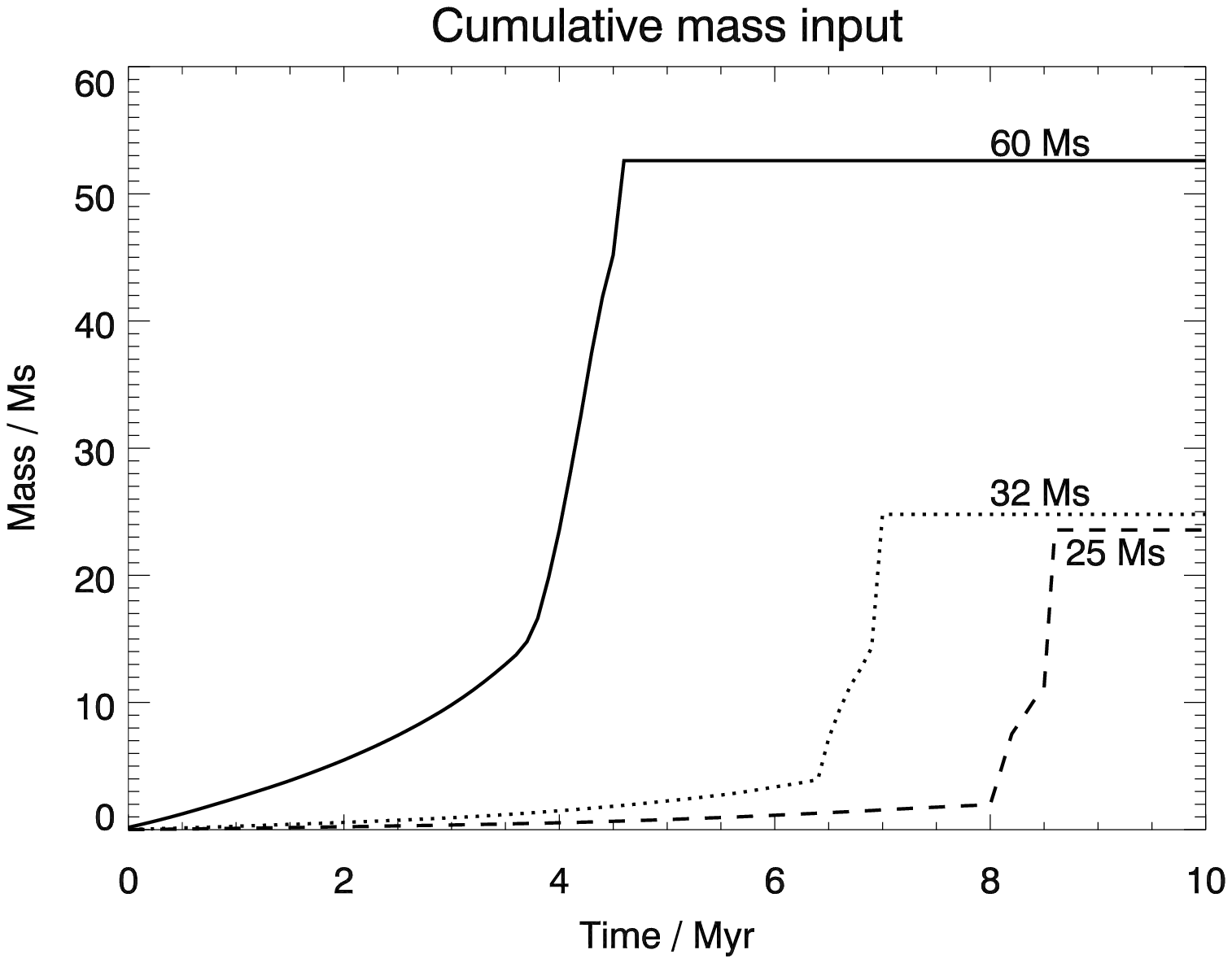}
  \caption{\rev{Cumulative} energy (top) and mass (bottom) input.  We use the output of
    a 25~$M_\odot$ (dashed, labels: Ms$=M_\odot$), 32~$M_\odot$
    (dotted) and a 60~$M_\odot$ star as input for our simulations,
    separately or combined. 
    The difference between the total mass output and the initial mass
    indicates the mass of the dark remnant.
    The energy is given in ``Bethe'' =
    $10^{51}$~erg.}
  \label{f:input}%
\end{figure}

The properties of the interstellar medium (ISM), i.e. its morphology
with imprinted bubbles and superbubbles
\citep{Gruendea00,Arthur07,Chu08,Sasea11} as well as molecular-cloud
fragments in formation or in dispersal, and its level of turbulence,
are strongly affected by the physics and dynamics of stellar feedback
\citep[e.g.][]{dAB04,dAB05,DBP11a,DBP11b,Ntormea11}.  \rem{Additionally, gas flows on larger scale,
  such as kpc scale gas streams driven by the gravity of the galaxy's
  dark- matter halo or the disk or the supermassive black hole, or
  smaller-scale gas streams resulting from gravity and imprinted ISM
  dynamics, might initiate and drive turbulence within a galaxy's
  interstellar medium.  } The actual agents of stellar feedback are
massive stars, born in the denser parts of the interstellar medium
\citep[for recent reviews see][]{MO07,Zy07}.  The interaction via
winds and ionising radiation of a single massive star with its
surroundings is usually referred to as ``interstellar bubble''
\citep{Weavea77}: Strong winds are shocked close to the star and
produce a hot overpressured bubble, which drives an expanding shell of
swept-up, shocked ambient gas. The shell may be partially or
completely ionised by the Ultraviolet emission of the central
star. The increased pressure may additionally push the leading shock
front.  Interstellar bubbles are usually not energy conserving,
because of the radiative losses of the shocked ambient medium
\citep[e.g.][]{Weavea77}. This has been nicely demonstrated \ch{by the
  observations of} \citet{Gruendea00}: For some bubbles, they resolve
the radiative leading shock wave, with the highly excited
\sqrbrk{O~III} tracing the hottest outermost gas, and H$\alpha$
tracing a somewhat cooler surface inside of \sqrbrk{O~III}. This
indicates that the leading shock front in these cases is shock ionised
rather than photo-ionised.  Radiative energy losses are substantial,
\ch{but hard to quantify in detail} \citep[e.g.][]{GSML95a} and affect
wind-blown and supernova related bubbles alike.
%
%
\begin{table*}
  \caption{Simulation parameters}             
  \label{t:simpars}      
  \centering          
  \begin{tabular}{l r r r r r r r }     
    \hline\hline       
    Label & Star mass /$M_\odot$ & X / pc & Y / pc & Z / pc & Res. / pc 
    & $n_0$/cm$^{-3}$ & $T_o /$~K\\
    \hline                    
    S25     &  25  &  0  &  0  &  0  &  2.1  & 10 & 121 \\
    S32     &  32  &  0  &  0  &  0  &  2.1  & 10 & 121\\
    S60     &  60  &  0  &  0  &  0  &  2.1  & 10 & 121\\
    3S0     &  25  &  0  &  0  &  0  &  2.1  & 10 & 121\\
    &  32  &  0  &  0  &  0  &   &  & \\
    &  60  &  0  &  0  &  0  &    & & \\
    3S1     &  25  &  -30  &  10  & 10  &  2.1  & 10 & 121\\
    &  32  &  -25  &  -10  &  0  &   & & \\
    &  60  &  0  &  0  &  0  &   &  & \\
    3S2     &  25  &  -60  &  20  &  10  &  2.1  & 10 & 121\\
    &  32  &  50  &  -10  &  0  &   & & \\
    &  60  &  0  &  0  &  0  &  &  & \\
    3S1-mr &  25  &  -30  &  10  & 10  &  1.0  & 10 & 121\\
    &  32  &  -25  &  10  &  0  &  & & \\
    &  60  &  0  &  0  &  0  &   & & \\
    3S1-hr &  25  &  -30  &  10  & 10  &  0.52  & 10 & 121\\
    &  32  &  -25  &  10  &  0  &  & & \\
    &  60  &  0  &  0  &  0  &   & & \\
    \hline                  
  \end{tabular}
\end{table*}

Many molecular clouds host massive stars in groups. The bubbles of
these stars have to interact, because the sizes of the individual
bubbles \citep[parsecs, e.g.][]{Weavea77,Gruendea00} is comparable to
the size of the parent molecular clouds \citep{Kainulea11}. Also, in
star forming regions, smaller groups of stars are often located within
distances of tens of parsecs \citep[e.g. Orion,][]{Vossea10}.  The
interaction of individual bubbles leads to the formation of
superbubbles (\citeauthor{TB88} \citeyear{TB88}; \citeauthor*{OCM01}
\citeyear{OCM01}; \citeauthor{Chu08} \citeyear{Chu08};
\citeauthor{Oey09} \citeyear{Oey09} for reviews): The expansion of the
combined superbubble is often described by the same model as for
individual bubbles, which predicts expansion rate and shell size, if
the energy input and the ambient density are known. Superbubbles may
reach sizes of hundreds of parsecs
\citep[e.g.][]{TB88,BdA06,Sasea11}. \ch{But they often appear to be}
too small and too bright in X-rays compared to models
\citep[e.g.][]{OG04,Jaskea11}.  Possible explanations include energy
dissipation due to mass loading or uncertainties in the stellar wind
data, e.g. due to clumping.  

Understanding of the physics of bubbles and superbubbles is the key
ingredient in order to gauge the efficiency of stellar feedback.  It
is of particular importance to assess the effective energy input into
the ISM. Our group has embarked on this task, and has synthesised the
total energy input into molecular clouds for realistic stellar
populations based on recent stellar evolution models \citep{Vossea09}:
\ch{Averaged over all massive stars ($8 M_\odot<M<120M_\odot$), the
  energy input due to winds is of order $10^{50}$~erg/star.}
Supernovae contribute about ten times more. The energy injection is
extended over several tens of Myr and has a peak near four Myr with a
shallow decline afterwards. Winds dominate before the peak and
supernovae afterwards. Substantial variations from cluster to cluster
are expected due to the sparser sampling at the massive end of the
initial mass function.

Stars are born in the densest regions of the ISM. Much of the injected
energy is therefore quickly lost to radiation in cooling shock
compressed shells. Hydrodynamic simulations have been used to assess
the effective energy input into the ISM.  The energy deposition
efficiency of isolated massive stars in their wind phase has been
assessed in 2D hydrodynamic simulations by \citet{FHY03,FHY06}. Though
they include the effect of photo-ionisation, they show that the gas
dynamical effects are dominated by the mechanical energy input: For
example, for a 35 (60)~$M_\odot$~star, they expect only 17 (5)~per
cent of the energy transfered to the ISM in their simulation being due
to the effect of ionisation. They give their energy deposition
efficiency as fractions of the radiative energy input. Scaled to the
mechanical energy input, they find that about 38 \ch{(9)}~per cent of the
input energy has been added to their ISM at the end of their
simulations for the 35~\ch{(60)}$M_\odot$ star.
The dynamics of two wind bubbles (25~$M_\odot$ and 40~$M_\odot$ stars)
separated by 16.2~pc has been studied by \citet{vanMea12}. The two
bubbles quickly merge, sweeping the colliding parts of the wind shells
away into the bubble of the lower mass star, due to the pressure
difference in the bubbles. An aspherical superbubble is then formed,
which isotropizes after a few Myr. More interesting details are
observed which we refer to below, when we compare them with our
findings in Section~\ref{res}. \citet{Ntormea11} have simulated the
merging of two superbubbles in 2D with identical stellar content.  One
of the most interesting findings in the simulations of
\citet{Ntormea11} and \citet{vanMea12} is the occurrence of the
Vishniac thin shell instability \citep{Vish83}.  This instability is
strongly suppressed in the simulations of \citet{FHY03,FHY06} due to
the thickening of the shell because of the increased pressure due to
the ionisation. The Vishniac instability is interesting as it may
create observable filamentary features, and thick filamentary shells,
and thus discriminate between models\revtwo{\citep{vMK12}}.

Here, we address the effective energy injection into a homogeneous ISM
for three interacting interstellar bubbles with 3D hydrodynamics
simulations, using standard ISM thermodynamics. We neglect the effect
of ionising radiation, because it is expected to be less important
\ch{in this context} (\citet{FHY03}, see e.g. \citet{Gritea10} for the
effects of ionising radiation).  We take as our starting element a
group of three coeval massive stars, 25, 32 and 60~$M_\odot$,
respectively.  We study the deposition efficiency of energy injection
as a function of distance between the stars. We find a high efficiency
in the wind phase, comparable to the 2D results of \citet{FHY03,FHY06}
\ch{and the 2D and 3D results of \citet{Fierlea12a},} details of
bubble merging similar to \citet{vanMea12} and an enhancement of the
feedback efficiency by about a factor of two for grouping of the stars
closer than about a few tens of pc. The energy of supernovae that
explode within superbubbles is dissipated on a timescale of about
1~Myr.  Additionally, we \revtwo{show 
column} density renderings of
prominently Vishniac unstable 3D shells, which should give a first
approximation of the observational appearance of the Vishniac
instability.

\section{Simulations}
We carry out 3D~hydrodynamic simulations with the \nv code
\citep{Z08,Z11}, evolving the conservation equations for mass,
momentum and energy.  \nv  is a conservative, finite volume code
and combines block structured adaptive mesh
refinement (AMR) with parallelisation by the \ch{message} passing
interface (MPI) library. 

\subsection{Numerics and code tests}
The main solver modules are an HLLD solver (HLLD\_CT), applying the
1D~approximate Riemann solver of \citet{MiKu05} dimension-by-dimension
in 3D, and a second-order Central-Upwind scheme \citep[CU\_CT, full
details in][]{Z11}. We work in Cartesian coordinates throughout \revtwo{(apart
from the radiative test case in this section)}. In
order to check the isotropy
of the solution in this geometry and also for differences between
these solvers, we have re-run and analysed the adiabatic blastwave
test problem that comes with the code with both solvers.
Here, a fixed amount of thermal energy is
initially deposited in a finite circular region of 22 cells
diameter. In both cases, a reasonably spherically symmetric bubble
develops, with a forward shock, a contact surface and a backward
shock. The contact surface evolves identically for both solvers,
forward and backward shock are lead by the solution of the CU\_CT
solver by at most one cell. Hence, both methods yield a very similar
result for symmetrical bubble expansion. For the same solver but 
different angular directions other than the grid axes, 
the radii of the different features of interest differ by
typically one and up to about three grid cells.

\revtwo{As a radiative (see below for details about radiative cooling and
  heating) test simulation, we have re-run the 1D-supernova test of
 \citet[Figure~\ref{f:TTtest}]{Tenea90}, also in spherical
 coordinates, but otherwise with the same numerical settings as for
 the 3D production runs below. Here, $10^{51}$~erg are
 deposited within a radius of $10^{18}$~cm. Density and temperature
 are initially assumed to be 1~cm$^{-3}$ and~100~K,
 respectively, throughout the computational domain. 
We use the CU\_CT solver
 with a uniform mesh with a cell size between $5.6\times 10^{-3}$~pc
 (6400 cells in total) and 0.36~pc (100 cells in total) for this
 test. The density slices in Figure~\ref{f:TTtest} (top) show the
 expected shape for such an explosion. In their 1D test run, 
 \citet{Tenea90} find the outer
 shock at 11~pc (24~pc) at 5,300~yr (47,806~yr). 
Our simulation (10.4~pc at 5483~yr;
 25.2~pc at 47,990~yr) reproduces this within expectations.  
At 47,990~yr, our outer shock is
 about 5~\% further out then their solution at 47,806~yr. 
This is likely related to the differences in the employed cooling
functions (more details below).
 The contact surface at 47,806~yr should be at 8~pc, which agrees well
with our result. During the energy conserving phase, i.e. up to say
30,000~yr, we expect 28~\% of the energy in kinetic form and 72~\% in
thermal form, which is consistent with our energy tracks
(Figure~\ref{f:TTtest}, middle, to be compared with Figure~1a in
\citet{Tenea90})
Cooling should become significant around 33,000~yr,
which is also in good agreement. After the onset of cooling, the
thermal energy should decline strongly, this and the shape of the
energy tracks are quite similar to the findings of \citet{Tenea90}. We
also find a secondary shock wave in the shocked ambient gas due to the
non-uniform cooling of the shell, and a corresponding increase in the
track of the kinetic energy, as in \citet{Tenea90} (another weak shock
from reflection at the origin is visible in the shocked ambient gas in
Figure~\ref{f:TTtest}, top).
The analytically expected compression ratio at the leading shock front is four.
Because of the strong decline of the solution inwards, one can however
not expect to obtain exactly four in a numerical representation, but
the solution should converge towards four with increasing
resolution. Our highest resolution run reaches a compression above~3.9
in the adiabatic phase, and we show in Figure~\ref{f:TTtest} (bottom)
that this value converges well with increasing resolution.

The reduced density peak height decreases the cooling rates slightly:
The low resolution runs lag behind in thermal energy decrease 
by at most about 3000~yr at 60,000--90,000~yr. The energy track is entirely
converged from about 400~cells.
The crossing point of thermal and kinetic energy is around 80,000~yr in
our simulation compared to about 46,000~yr in \citet{Tenea90}. This
significant difference is due to the employed cooling curve: 
\citet{Tenea90} use the cooling curve of \citet{RCS76}, which features
particularly strong cooling around $10^6-10^7$~K due to highly ionised
Fe. The cooling rates
are uncertain by a factor of about two \citep{WSS09}. Many, more recent,
cooling curves tend to have lower cooling rates than \citet{RCS76},
including the standard one for the NIRVANA-code, \citet{Slyzea05},
which we use. In order to verify that this is the reason for the
differences in the energy tracks between \citet{Tenea90} and our
result, we tested a case where we increased the
cooling rates ad hoc by a factor of two (Figure~\ref{f:TTtest},
middle, blue dot-dashed lines). This obviously shifts the result into
the right direction. The
increased amount of thermal energy probably also leads to the slightly
further advanced outer shock (compare above) at late times. 

Thus, we reproduce the fundamental properties of the 1D test of
\citet{Tenea90}. The 3D nature of our simulations demands some
compromise regarding resolution. 
We expect that
this effect should affect energy tracks by a at most a few per cent. We
account for this
in the discussion below.}

We have initially selected the HLLD\_CT solver but encountered severe
vacuum formation problems (very low pressure) near contact surfaces for our high
resolution runs. For all the simulations presented in this article, we
have therefore employed CU\_CT.

We use standard ISM thermodynamics with radiative cooling and
photo-electric heating (see \citet{PGZ09} for details), employing the
standard iterative procedure of \nve.  Cooling is always strong for
our wind shells, which tend to get thin and eventually also
Rayleigh-Taylor and Vishniac-unstable. The instabilities evolve
differently for different flux limiters: Test simulations
with all flux limiters provided (minmod, superbee,
monotonised-centred, and Van Leer) showed that for the
monotonised-centred and the superbee limiters, the
instabilities are systematically different for parts of the shell
which move parallel and diagonal to the grid axis. Van Leer and minmod
both yield almost isotropic results at our highest resolution, at the
expense of being more diffusive, as expected. We have
correspondingly adopted the minmod flux limiter.  \nv offers an
additional multi-dimensional limiter which we also use, and where we
have adjusted the parameter experimentally to yield optimal isotropy
for shell instabilities.

%
\begin{figure*}
  \centering
  \includegraphics[width=0.9\textwidth]{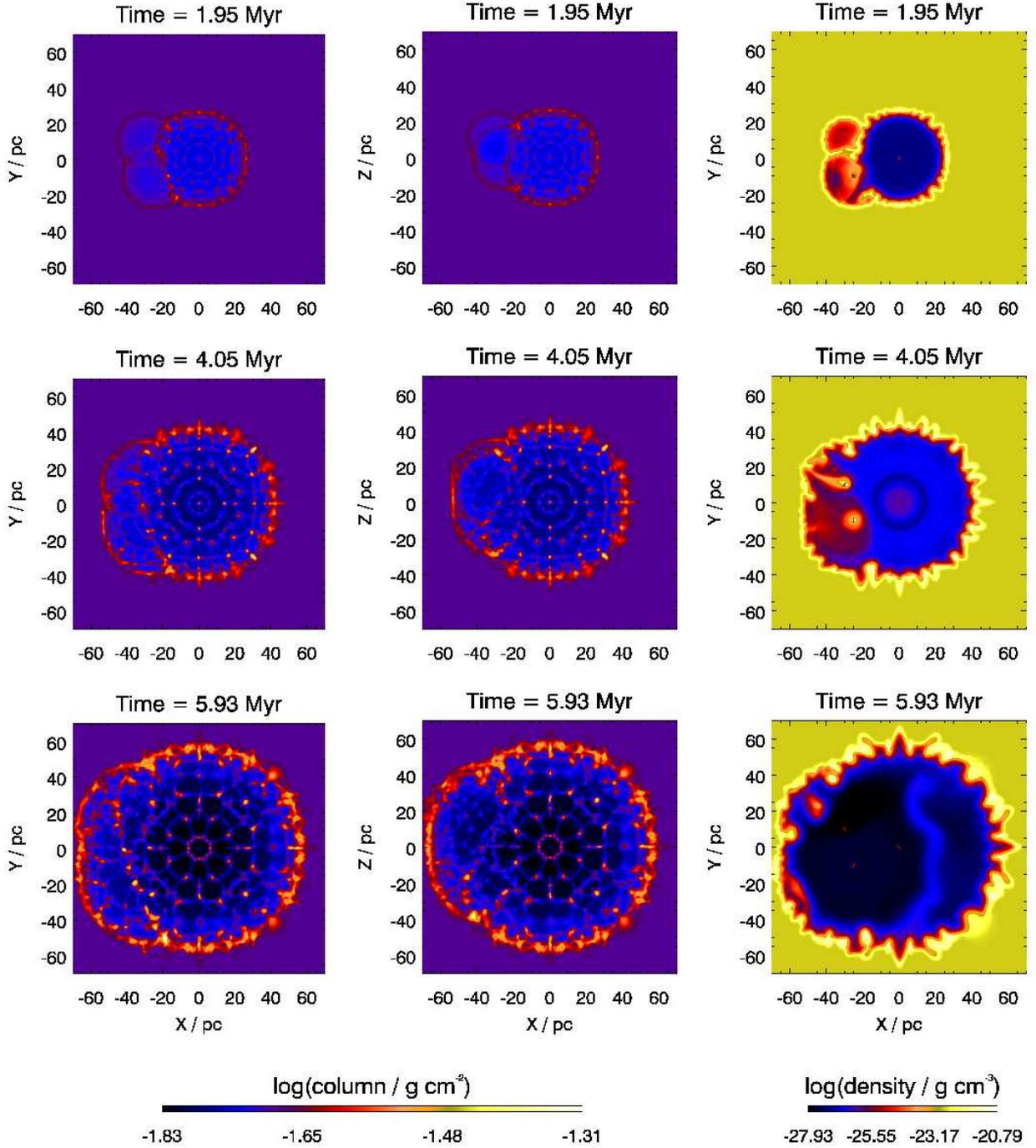}
  \caption{Column density integrated over $Z$-direction and
    $Y$-direction (left and middle columns, respectively) and midplane
    density (right column) for three different snapshot times from top
    to bottom for run 3S1-hr. The projections of the three massive stars into
    the X-Y plane is indicated as small red stars in the  density plots on
    the right. The 60~\ms star blows the biggest bubble
    from the origin. The 32~\ms bubble towards its lower left
    ($XY$-plots) is only slightly bigger than the one of the 25~\ms
    star above. The shell forms spikes and dense clumps due to the
    combined action of Vishniac and thermal instability. \ch{A movie is
    provided with the online version.}}
  \label{f:3sw1-overview}%
\end{figure*}
\begin{figure*}
  \centering
  \includegraphics[width=0.9\textwidth]{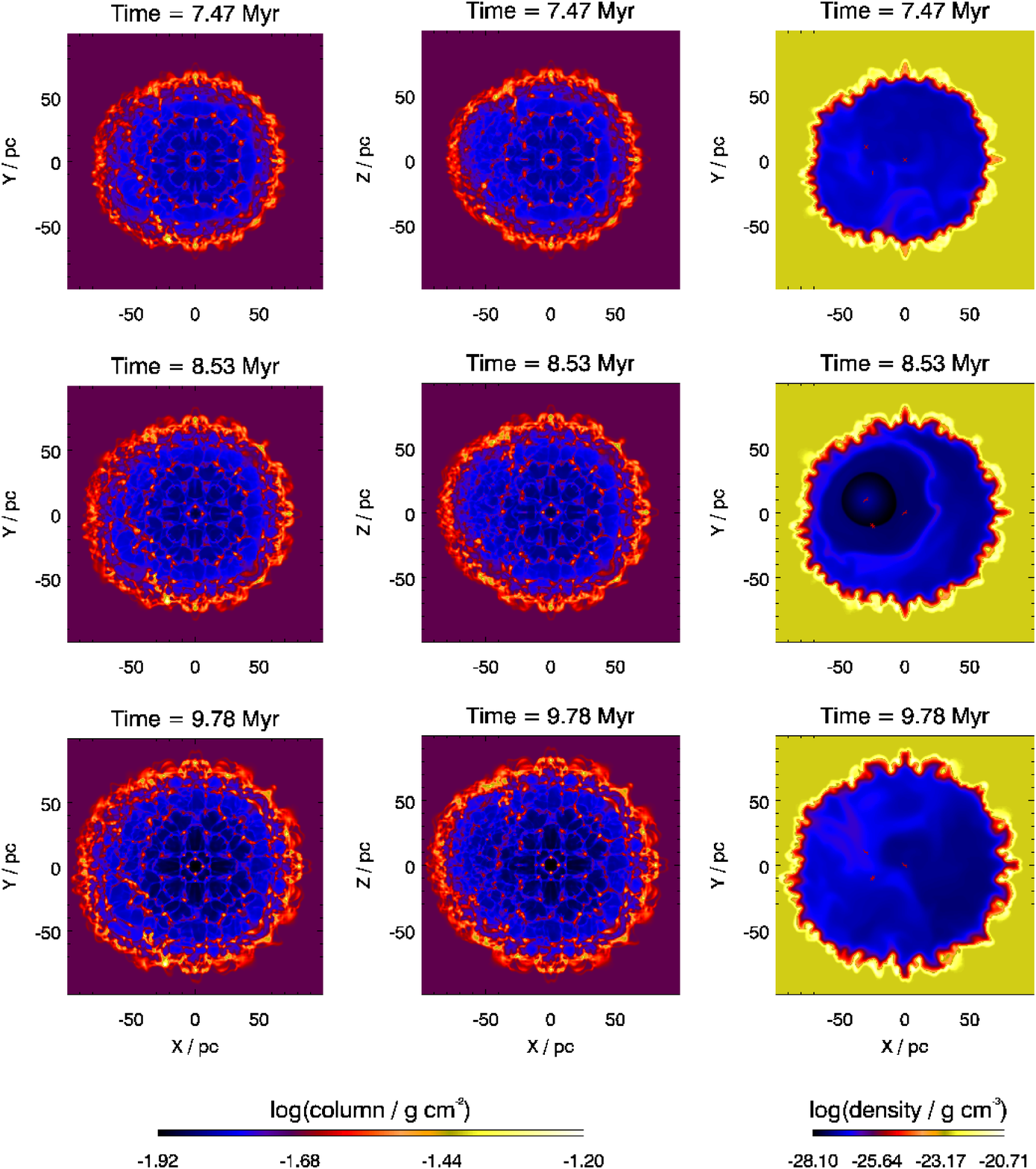}
  \caption{Figure~\ref{f:3sw1-overview} continued, but with all scales
    adapted to the snapshots presented in this figure.}
  \label{f:3sw1-overviewb}%
\end{figure*}
%
%

\subsection{Setup}

The computational domain is a cubic Cartesian grid, 400~pc on a side
resolved by 24~cells for the base level. The mesh is refined whenever
a combined threshold of first and second derivative for density or
respectively velocity is exceeded. Additionally, we always keep the
wind injection region at the highest refinement level. Effectively,
the wind shell and everything inside is always refined to the highest
level.  For most of our runs we use three levels of adaptive mesh
refinement, which would correspond to a uniform grid of $192^3$~cells
with a resolution of 2.1~pc. Simulation 3S1-mr and 3S1-hr use four and
five levels of refinement, resulting in 1 and 0.5~pc resolution,
respectively.  Boundary conditions are formally periodic, but we only use
data from snapshots where the shells are entirely contained in
the computational domain.

We fill the grid initially with a homogeneous medium.  \ch{Then we
  choose one (three) injection regions \ch{of eight pc radius in every
  case}. Each injection region gets assigned a star of a particular
mass.
We inject mass and thermal
  energy according to the stellar evolutionary tracks of rotating
  stars of \citet{MM05} and wind velocities from \citet{Lamea95} and
  \citet{NiSk02} for the Wolf-Rayet phase, as compiled in
  \citet{Vossea09}.}  We use 25, 32, and~$60 M_\odot$ stars, with
supernovae at 8.6, 7.0 and~4.6~Myr, respectively.  The time resolution
of the stellar evolution table is 0.1~Myr.
\rev{Cumulative} mass and energy input are shown in Figure~\ref{f:input}.  The mass
density is initially set to 10~$m_\mathrm{p} $~cm$^{-3}$ everywhere in the
computational domain. The temperature is set in equilibrium
between cooling and heating, 121~K. 
All velocities are initially zero. More details for each
individual run are provided in Table~\ref{t:simpars}.

%
%
\begin{figure}
  \centering
  \includegraphics[width=0.49\textwidth]{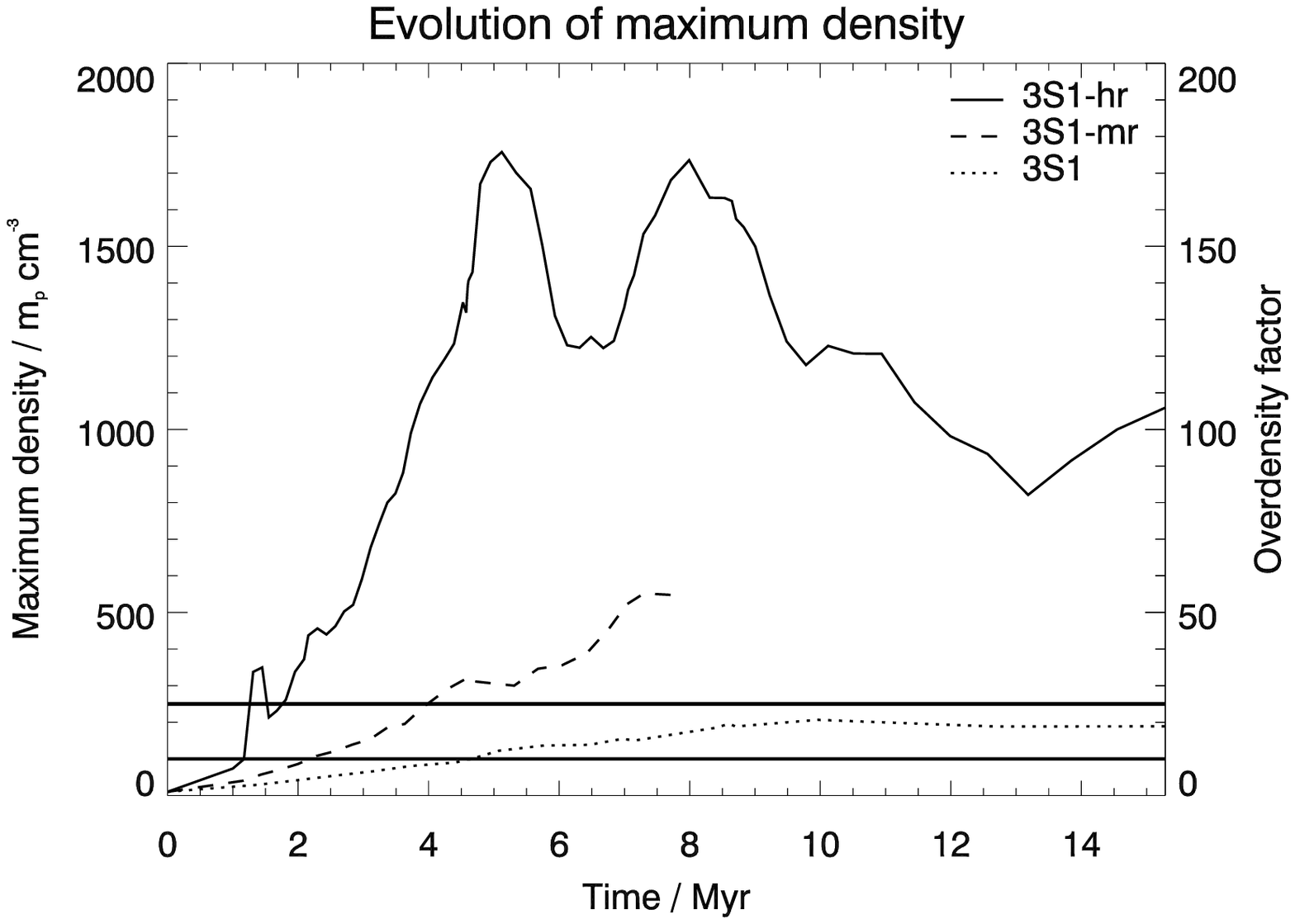}
  \caption{Maximum density as a function of time for runs
    3S1-hr (solid), 3S1-mr (dashed) and 3S1 (dotted).  The horizontal
    lines correspond to the critical compression above which the
    Vishniac instability is triggered for supernova (lower line) and
    wind (thicker upper line) shells according to \citet{VR89}.  The
    axis on the right shows the overdensity factor over the
    undisturbed ambient medium.}
  \label{f:3sw1_maxd-t}%
\end{figure}
%
%
%
%
\begin{figure}
  \centering
  \includegraphics[width=0.49\textwidth]{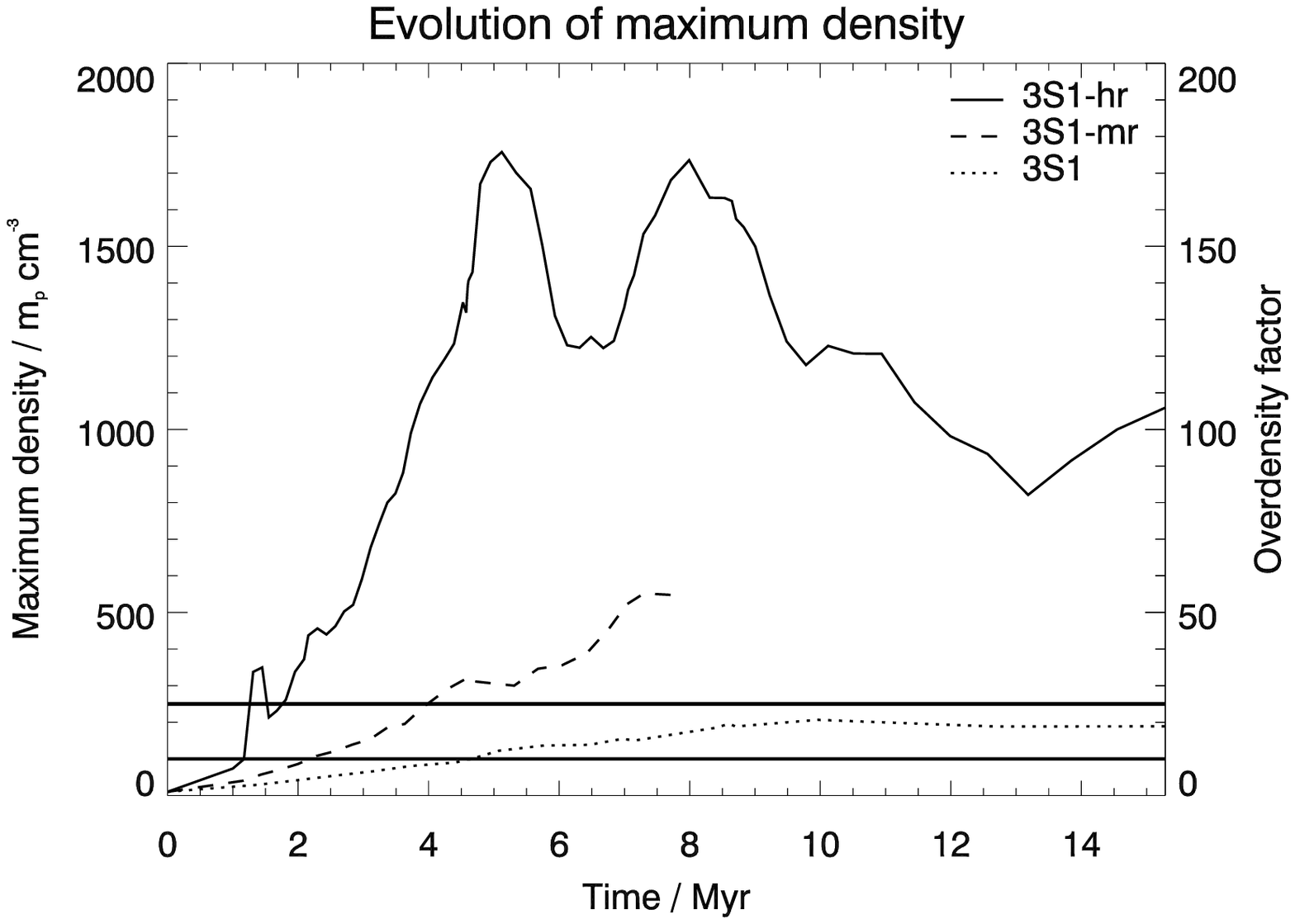}
  \caption{One-dimensional slices in X-direction through run 3S1-hr at
    time $T=8.53$~Myr. The $Y$ and $Z$ coordinates are chosen
    appropriately for the slices to include the position of the only
    remaining star at that time
    (25~\ms, at $X=-30$~pc, indicated by the star in the middle
    diagram). 
    Top: positive X-velocity (blue, dashed line), negative
    x-velocity (red dash-dotted line) and sound speed (solid
    black). Middle: pressure (logarithmic) . Bottom: density
    (logarithmic). See text for details.  }
  \label{f:3s1-1d}%
\end{figure}
%
%
%
%
\begin{figure}
  \centering
  \includegraphics[width=0.49\textwidth]{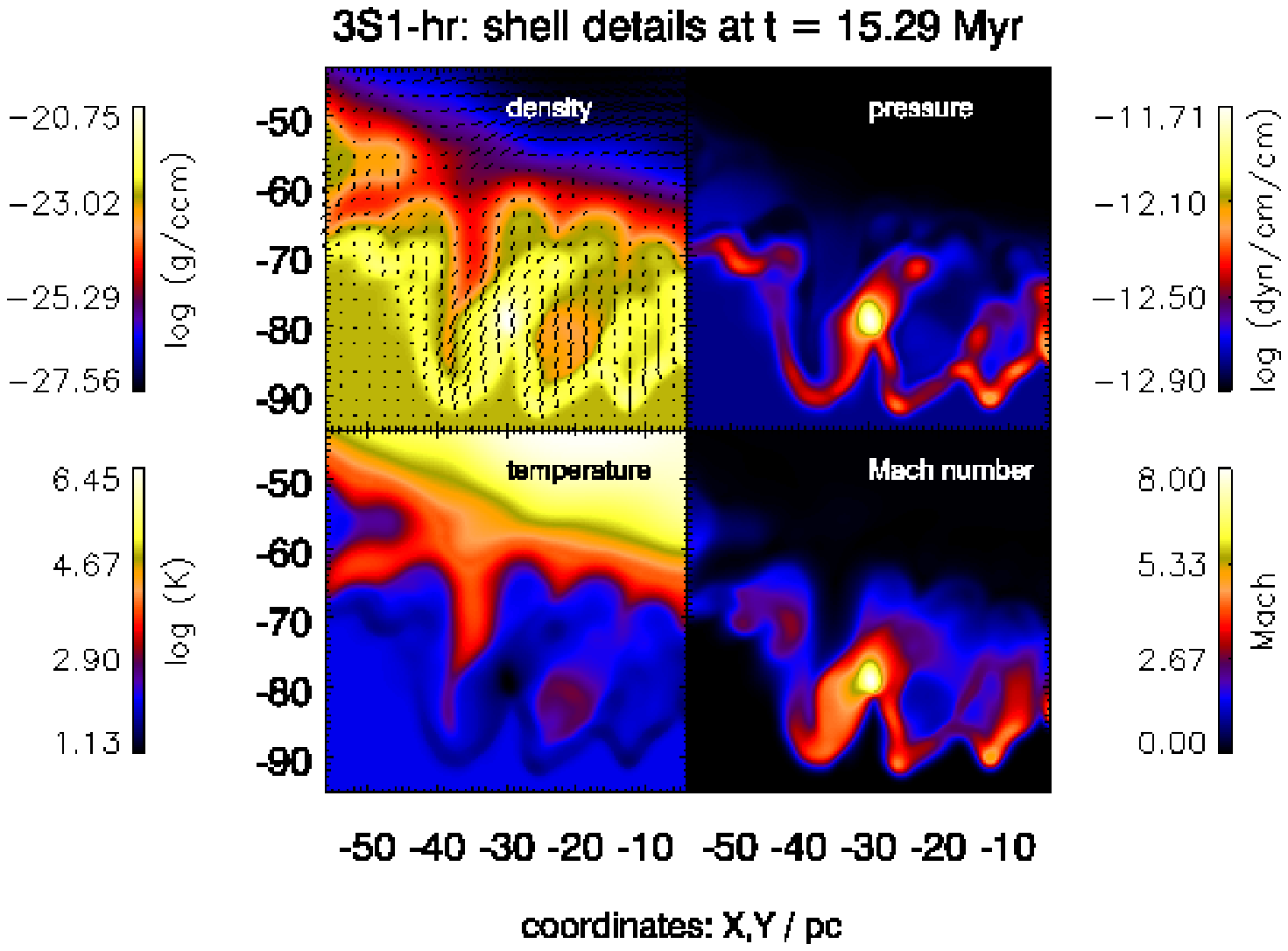}
  \caption{Shell details for the final snapshot of run~3S1-hr. Shown is
    an X-Y~zoom of density, pressure, temperature and Mach number, as
    indicated on the individual panels, around the position of the
    maximum density, which is located at
    $(X,Y,Z)=(-30,-79,-52)$~pc. Velocity vectors are overlaid on the
    density plot. The high density region is overpressured and has a
    temperature below 20~K. See text for more details.  }
  \label{f:3s1-hdzoom}%
\end{figure}
%
%

\section{Results}\label{res}                                  
The time evolution of our high resolution run 3S1-hr with three stars
at different locations is shown in
Figures~\ref{f:3sw1-overview}~and~\ref{f:3sw1-overviewb}.  At a given
time, the bubble size increases monotonically with the mass of the
parent star, with the central 60~\ms bubble dominating the gas
dynamics. As expected, the shocked ambient medium cools very quickly
and consequently gets compressed into a thin shell for each
bubble. The shell is subject to a combination of thermal and
\citet{Vish83} instabilities\ch{\footnote{Although we have carefully chosen the
flux limiter, the shell instability evolves still somewhat
anisotropically. This is similar to the 2D results of
\citet{Ntormea11} with the RAMSES code, where even a five times higher
spatial resolution could not get the shell instabilities completely
isotropic.}}. The bubbles start to merge at around 2~Myr. At the first
snapshot \ch{in Figure~\ref{f:3sw1-overview}} (1.95~Myr), the shell
interface between the 60~\ms bubble and the 32~\ms bubble has just
burst. Up to this point, each bubble has had its individual bubble
pressure, which is largest for the 60~\ms bubble. Its hot gas can be
seen to stream through the hole in the shell. The shell interface then
behaves much like a cloud, being ablated by a wind \citep{Pitea05}:
Kelvin-Helmholtz instabilities at the contact surface lead to mixing
of the cloud gas into the hot phase. The shell interface has
completely dispersed until the next snapshot at 4.05~Myr.  We have
checked the effect of different flux limiters in this phase: Less
diffusive ones allow smaller holes, which delays the erosion process
compared to the more diffusive case. The final results are however
very similar.

The density slice at 4.05~Myr shows the weaker winds of the smaller
stars to be pushed aside by the one of the most massive star. The
larger part of the 60~\ms bubble remains unaffected by the action of
the smaller stars. The 60~\ms star explodes at 4.6~Myr. The sudden
energy injection due to the supernova compresses the shell further
(Figure~\ref{f:3sw1_maxd-t}) and accelerates it, triggering the
Rayleigh-Taylor instability (RTI). The RTI may cause filamentary
structure inside the shell. Also, the outwards directed flow field, 
centred around 
the most massive star before its explosion, 
is no longer present. Thus, from this time on,
we find filamentary gas inside the shell, seen in the individual
density slices. The effect of the winds of the smaller stars in this
phase can hardly be noticed.  The second supernova (7.0~Myr) leads to
a further acceleration and compression of the shell, causing more RTI
filaments. The snapshot at 8.53~Myr shows the superbubble when 2 stars
have exploded already, and the third is in its Wolf-Rayet phase. This
snapshot demonstrates nicely that our ansatz with thermal energy
injection may also cope with situations when the backward shock within
the stellar ejecta is unusually far from the star: 
One can clearly see the declining density away from the star due
to adiabatic expansion (1D-slices in Figure~\ref{f:3s1-1d}). The wind
turns supersonic immediately outside the driver and shocks roughly
20~pc away from the star. A second structure is visible at varying
distance from the star, up to about 50~pc: This is what we would
expect to be the forward shock in the standard picture. Due to the
high ambient pressure, it is only a sound wave. The pressure inside of
this structure is slightly reduced due to the ongoing expansion. The
final supernova at 8.6~Myr causes again mass entrainment into the
bubble due to the RTI. The bubble then keeps expanding with decreasing
interior density fluctuations until the end of the simulation at
15~Myr.

The highest densities in the shell, around 180 times the ambient
density, are reached for roughly 1~Myr after each supernova, where for
the later two supernovae, the compression peaks have merged
(Figure~\ref{f:3sw1_maxd-t}). At late times the density increases
again (see below for details). 
We show a zoom on the highest density region in the final
snapshot in Figure~\ref{f:3s1-hdzoom}. The density maximum is located
in the dense shell, where two humps of the Vishniac instability
(compare Section~\ref{s:vish} below) cross, and more towards the
interior of the bubble. The velocity field in the shell is still
dominantly outwards with substantial Mach numbers.  Yet, probably
enhanced by the large scale vortices which dominate the shell interior
at that time, there is also some non-radial motion. The slightly
converging velocity field has to be responsible for the high density,
as the region is substantially overpressured compared to the
environment.  At earlier times (compare above), such maxima in density
and pressure could have been in pressure equilibrium with their
surroundings. At this late time, the bubble interior is already underpressured with
respect to the environment, and so we expect that the
maximum is temporary, unless such clumps become self-gravitating. This
seems quite likely, given the pc-scale size, low temperature (below
20~K) and high mass (few hundred \ms) of the clump (Jeans length:
$\approx 2$~pc). Yet, self-gravity
is not included in the simulations and therefore details, such as
triggered star formation, are beyond the scope of this article.
%
%
\begin{figure}
  \centering
  \includegraphics[width=0.49\textwidth]{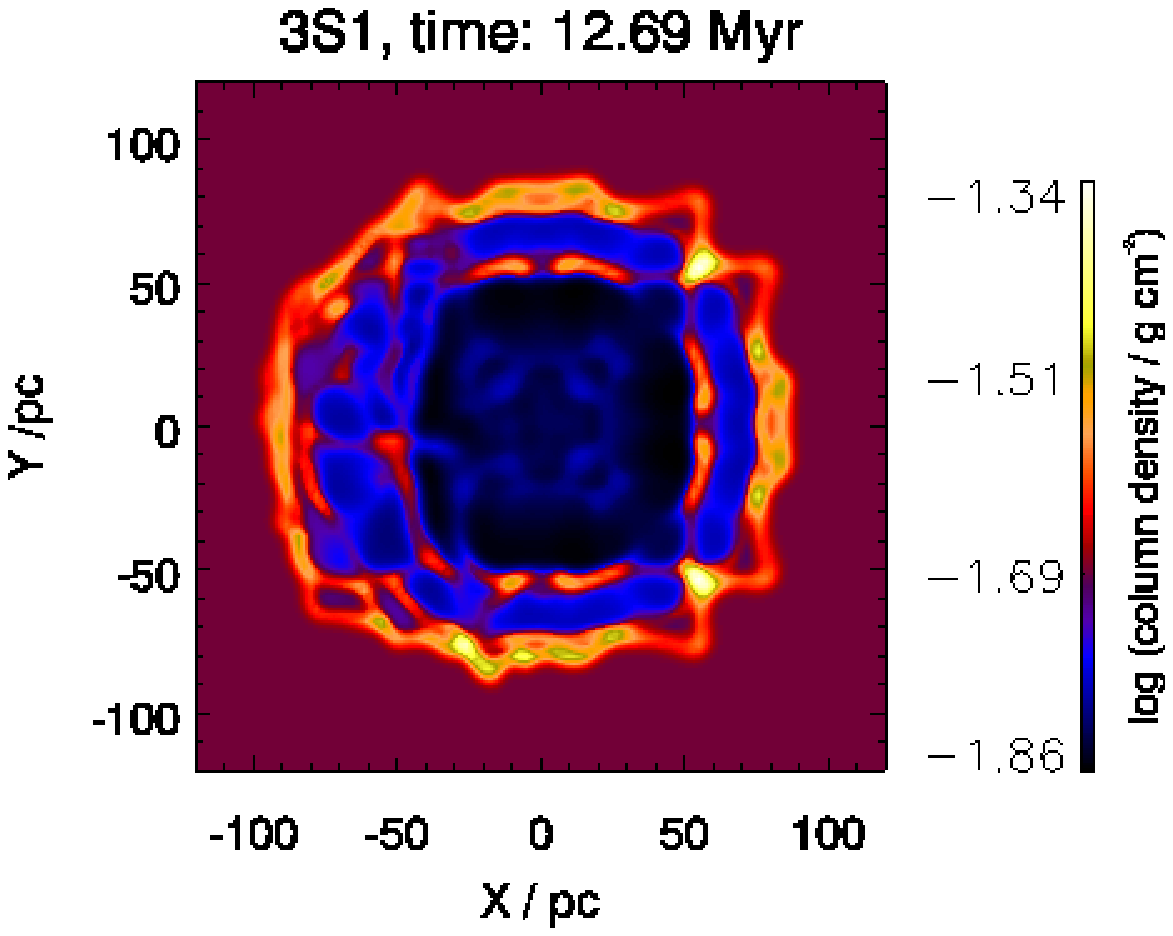}\\
  \includegraphics[width=0.49\textwidth]{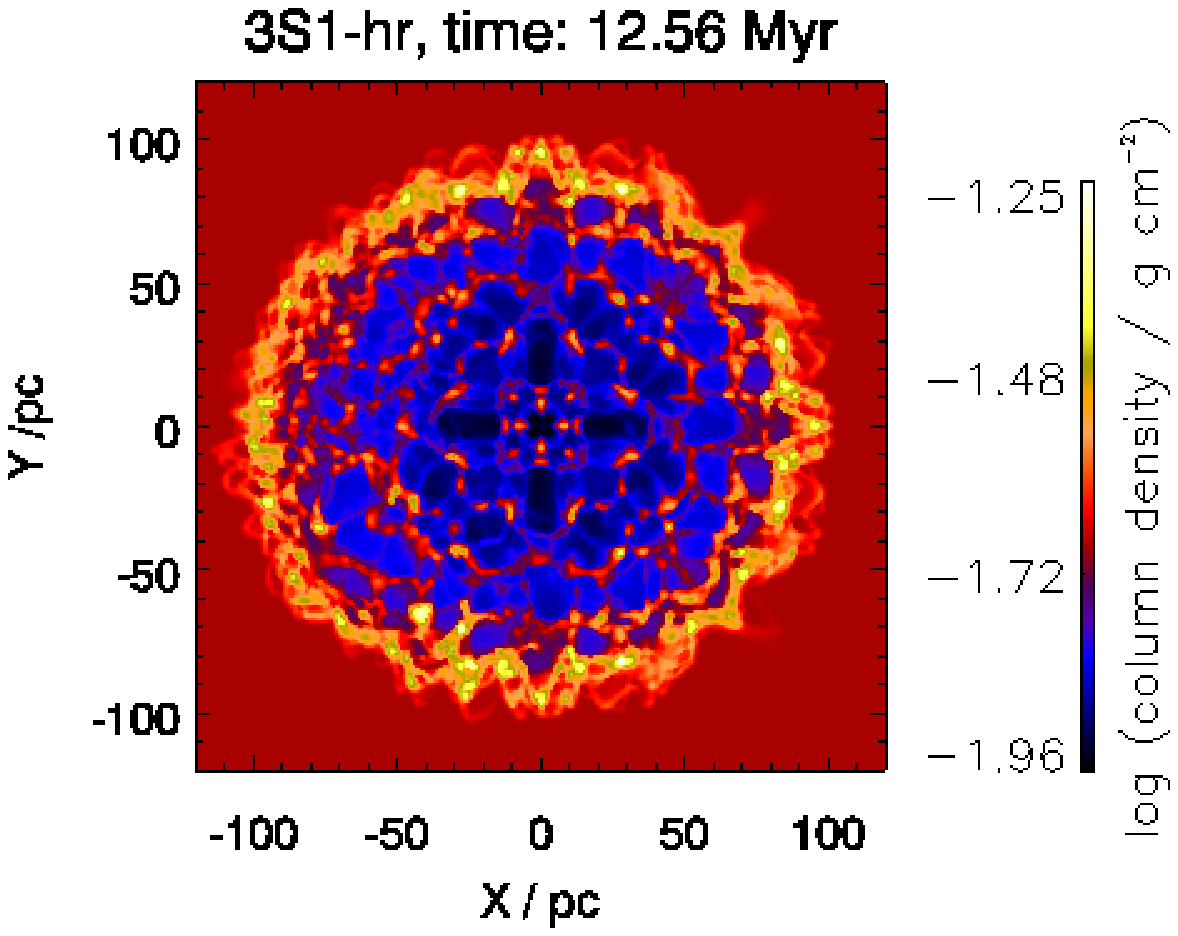}
  \caption{Column density at a comparable late evolution time for runs
    3S1 (top) and 3S1-hr (bottom). The high resolution bubble is more
    spherical, larger, achieves higher peak column densities and the
    Vishniac instability is more pronounced.  }
  \label{f:3s1_cd_rescomp}%
\end{figure}
%
%
%
%
\begin{figure}
  \centering
  \includegraphics[width=0.5\textwidth]{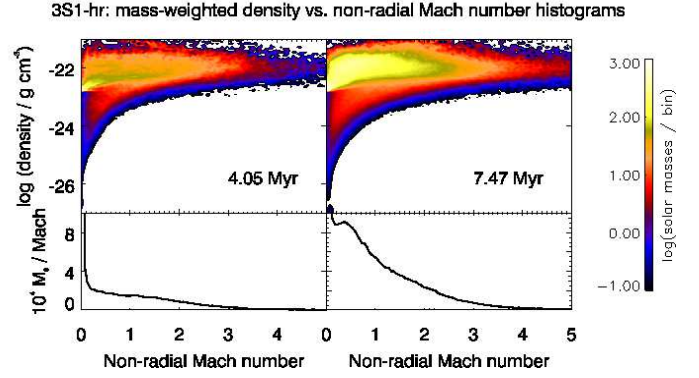}
  \caption{Analysis of the non-radial Mach number, i.e. the Mach
    number perpendicular to the direction of the shell's
    expansion. Only the region with positive $X$-coordinate, which
    corresponds to the undisturbed part of the 60~\ms bubble, is taken
    into account. Left: 4.05~Myr, right: 7.47~Myr. The upper parts
    show the non-radial Mach number versus the logarithm of the
    density. Colour encodes the mass per bin, where each bin spans
    0.05 in Mach number and 0.06~dex in logarithmic density. No
    appreciable non-radial motions are found for the hot bubble
    interior, whereas the dense shell material shows Mach numbers of
    order unity. The lower parts show mass weighted non-radial Mach
    number histograms (vertically collapsed versions of the plots
    above). The plots are dominated by the quiescent ambient
    medium. The mass with given non-radial Mach number declines
    strongly around a Mach number of unity towards higher Mach numbers
    as expected for shells dominated by the Vishniac instability.  }
  \label{f:3sw1-nonrMach}%
\end{figure}
%
%

\subsection{Vishniac instability}\label{s:vish}
\ch{The shells are subject to various instabilities. The
  Rayleigh-Taylor instability is especially prominent during the
  strong acceleration phases after each supernova.}  
\revtwo{The
Vishniac instability develops when the shell decelerates.
It} is an overstability: differences in column
density for adjacent regions of a shell cause gas flow from the high
column density region into the region with smaller column density. This
continues in general until the situation is reversed and the region
with initially smaller column density finally has the greater
one. \citet{VR89} derive a critical overdensity for the shell over the
unshocked ambient gas of a factor 10 and 25 for a blastwave with
initial energy injection and constant energy injection rate,
respectively, to become unstable, such that the peak density increases
in each cycle. The shell then develops a
characteristic spiky pattern
\citep[Figure~\ref{f:3sw1-overview}]{Ntormea11,Drake12}, in density
slices.  The 3D structure of the shell is granular with a regular
filamentary pattern
(Figure~\ref{f:3sw1-overview}~and~\ref{f:3sw1-overviewb}). The
regularity is of course related to the grid structure, because this is
the most important perturbation. In column density, we find a web
formed of polygons. These polygons have typically four to six
sides. The sides are however not always aligned with the coordinate
axis or the diagonals, and some are clearly curved. The typical
polygon diameter is about 10~pc. At the intersections of the
filaments, density and column density achieve their highest
values. These points lag behind the shell. Particularly high densities
may be achieved, when left and right part of an inwards spike
merge. This seems to have happened for the density maximum at the
final snapshot we show in Figure~\ref{f:3s1-hdzoom}. But from a
detailed inspection of several snapshots, we conclude that this should
happen frequently.
\revtwo{The three-dimensional structure of our superbubble shells is
  very similar to the one of the smaller scale circum-stellar shells
  of \citet{vMK12}.}

We show the peak density over time in
Figure~\ref{f:3sw1_maxd-t}. Clearly, the densest parts of the shell of
run 3S1-hr satisfy the criteria of \citet{VR89} from before 2~Myr
throughout the simulation, in agreement with
Figure~\ref{f:3sw1-overview}.  The low
resolution simulation 3S1 generally stays below the wind criterion of
\citet{VR89}. Correspondingly, the Vishniac instability is much less
pronounced (Figure~\ref{f:3s1_cd_rescomp}).  \citet{MLN93} have shown
that the instability is connected to transonic motions in the shell
perpendicular to the expansion direction. We evaluate these non-radial
velocities for the undisturbed (with respect to the interaction of the
bubbles of the other two stars, here we use $X>0$) part of the shell in
Figure~\ref{f:3sw1-nonrMach}. The 2D~mass weighted histogram over
logarithmic density and non-radial Mach number, with respect to the
local speed of sound, shows that only dense
shell gas acquires substantial non-radial Mach numbers.  At high
densities, indeed most of the gas has Mach numbers around and below
unity.

\subsection{Energy evolution: general observations}
\begin{figure}
  \centering
  \includegraphics[width=0.45\textwidth]{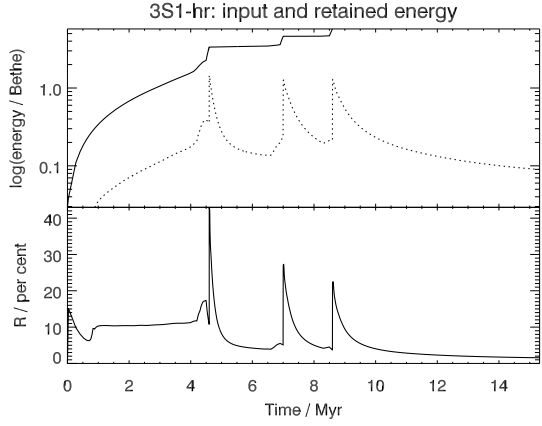}
  \caption{Top part: input (solid) and retained (dotted) energy for
    run~3S1-hr. The response $\cal R$ (retained energy divided by
    input energy) is shown in the bottom part. See text for more
    details.  }
  \label{f:3S1-hr-e}%
\end{figure}

We show the total input energy over time together with the energy
retained in the ISM where the initial thermal energy is subtracted in
Figure~\ref{f:3S1-hr-e}.  The retained energy is generally below the
input energy because the gas is initially in radiative equilibrium and
suffers net radiative losses during the course of the simulation. We
define the response $\cal R$ to be the energy retained in the ISM
divided by the input energy: \eql{eq:eres}{ {\cal R}(t) =
  \frac{E_\mathrm{ISM}(t)-E_\mathrm{ISM,0}}{E_\mathrm{in}(t)} \, }
where we define as ISM the whole gas present in the computational
domain, including the hot bubble interiors with their stellar ejecta.

$\cal R$ is generally of order ten per cent. It is higher whenever the
energy input rate increases. This is especially well visible at the
time of the three supernovae at 4.6, 7.0 and 8.6~Myr. Here, $\cal R$
reaches peak values between 20 to 40~per~cent. $\cal R$ is smaller for
phases of decreasing energy input rate. This is particularly well
visible after a supernova. About 1~Myr after each supernova, $\cal R$
drops to roughly five per cent. The characteristic decay time of the
retained energy increases for each consecutive supernova. When the
energy input ceases, the ISM energy is lost to radiation on a
timescale of Myr, with $\cal R$ dropping to 2~per~cent roughly 4~Myr
after the last supernova.

Steady, continuous energy injection is clearly more effective in
energising the ISM than sudden bursts such as from infrequent
supernovae.
%
%
%
%
%
%
\begin{figure}
  \centering
  \includegraphics[width=0.45\textwidth]{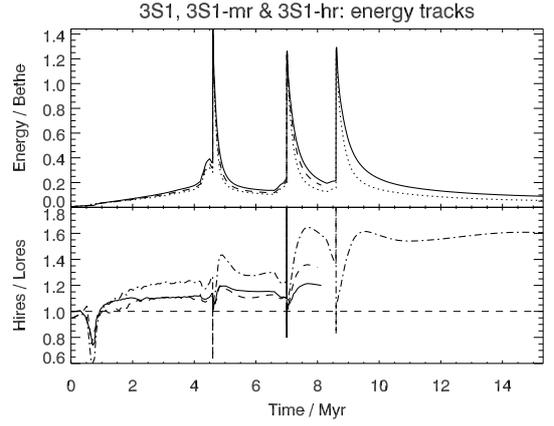}
  \caption{Resolution effects on the retained energy. Top part:
    retained energy for run~3S1-hr (solid line, high resolution),
    3S-mr (dashed line, intermediate resolution) and run~3S1 (dotted
    line, low resolution). The bottom part shows the energy ratio
    3S1-hr/3S1-mr (solid line), 3S1-mr/3S1 (dashed line) and
    3S1-hr/3S1 (dash-dotted line). In each case, the data for the
    higher resolution run has been interpolated to the data output
    times of the lower resolution run. The spikes at the supernova
    times are artefacts of the interpolation process at the
    discontinuities of the functions.  The horizontal dashed line
    indicates equality for comparison. The energy increases similarly
    for each doubling of resolution. The general functional behaviour
    is independent of resolution. See text for more details.  }
  \label{f:3sw1-eres}%
\end{figure}
%
%

\subsection{Resolution effects}
We have repeated run~3S1-hr at a half and a quarter of the original
spatial resolution. Morphologically, the bubbles are less spherical,
smaller and the Vishniac instability is less developed at lower
resolution (Figure~\ref{f:3s1_cd_rescomp}).  We compare the energy
evolution of the three runs in Figure~\ref{f:3sw1-eres}. The retained
energy differs by much less than a factor of two between simulations at
different resolution.  The differences are more pronounced at later
simulation times. Finer spatial resolution always leads to more energy
in the ISM. For an increase of the resolution by a factor of two, we
find an increase of the retained energy by 20-30~per
cent. This agrees with the greater bubble diameter at higher
resolution (Figure~\ref{f:3s1_cd_rescomp}).  The overall functional
behaviour is very well converged.

The reason for the changes with resolution is very likely the details
of the shell evolution. In the absence of other perturbations,
instabilities are triggered on the resolution level. Additionally, the
Vishniac instability is only marginally developed at low resolution.
This might lead to more non-radial kinetic energy at higher
resolution, which is not immediately radiated away. Also, the peak
density at a given time depends strongly on resolution
(Figure~\ref{f:3sw1_maxd-t}), which also changes the thermodynamics.

\subsection{Energy evolution: varying stellar distances}
\begin{figure}
  \centering
  \includegraphics[width=0.45\textwidth]{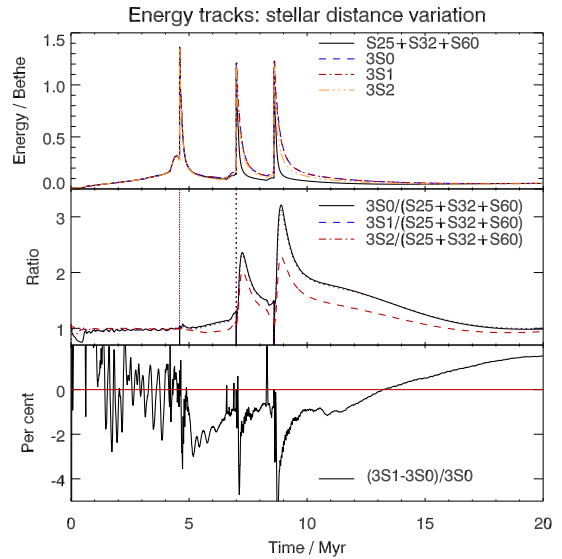}
\caption{Energy tracks for different simulations, where only the
  positions of the three stars differ. Run labels in the legends are
  explained in Table~\ref{t:simpars}. \mbox{S25+S32+S60} refers to the
  sum of the energy tracks of the three simulations of the
  bubbles of the isolated
  25~\ms, 32~\ms and 60~\ms stars, respectively, which corresponds to
  a very large distance. 3S0 is the opposite case, where three
  stars are in the same region. Top: Absolute values. 
Middle: Three-stars simulations relative to the sum of the
  three isolated bubbles. Interpolations always use the 3S0 time
  base. Interpolation artefacts are visible at the discontinuities due
  to the supernovae (4.6, 7.0 and 8.6~Myr). 
  Bottom: Difference of the very similar energy tracks of runs 3S1 and
  3S0, normalised to 3S0 as a percentage. The solid red line marks
  zero. The difference has been set to zero for the time intervals
  10,000~yr around each supernova in order to mask the interpolation
  artefacts. See text for details.  }
\label{f:etracks-dist}%
\end{figure}

We have carried out a set of simulations, where we varied the
positions and distances of the same three stars
(Figure~\ref{f:etracks-dist}). Because of computational limitations,
these simulations have been carried out at 2.1~pc resolution. This is
physically justified by the convergence of the general shape of the
energy tracks (Figure~\ref{f:3sw1-eres}). For obtaining the large
distance limiting case, we have simulated each of the three bubbles in
a separate simulation (S25, S32 and S60), and added their energy
tracks for comparison to the other cases. We model the closely-spaced
extreme case, where the bubbles have merged instantaneously, by
putting the driver regions of the three stars on top of each other at
the grid origin (3S0). Additionally, we performed two simulations with
intermediate star positions (compare Table~\ref{t:simpars}), where we
actually observe the bubble merging during the simulations (3S1 and
3S2).

We see small differences in the energy tracks during the first
$\approx 0.5$~Myr. They are expected because during this time, the
driver region is evacuated and the bubble shape is established. It
makes of course a difference, if the three stars share the same driver
region (3S0), or if each star has its own. Also shifting the driver
region on the grid makes the volume of the individual driver regions
slightly different, by a few per cent, due to resolution effects at
the driver boundary. This translates to a few per cent difference in
total energy, which is visible in Figure~\ref{f:etracks-dist}
(bottom).

Once the bubbles are established properly on the grid, i.e. after
about 0.5~Myr, all configurations have essentially the same energy
response until the first supernova at 4.6~Myr. The reason for this is
the predominance of the energy injection of the 60~\ms star.  The
energy tracks begin to differ slightly after the first star has
exploded. The divergence increases abruptly after each supernova. But
for very long times after the final explosion, the tracks converge
again towards a common value.

Among the four configurations, the energy varies \ch{at times} by up to a factor
of~\ch{three}.
A typical value \ch{after the second supernova} is a factor of two.  
Throughout the simulation time,
the energy is essentially highest for run~3S0 (all stars at same place) and lowest
for very large distance (sum of S25, S32 and S60). The two
configurations with intermediate distances, where the bubbles merge
during the respective simulations, show intermediate energies. The run
where the bubbles merge early (3S1) behaves almost identical to the
case where the driver regions are on top of each other (3S0).

%
%
%
%
\subsection{Shell widths}
\begin{figure}
  \centering
  \includegraphics[width=0.45\textwidth]{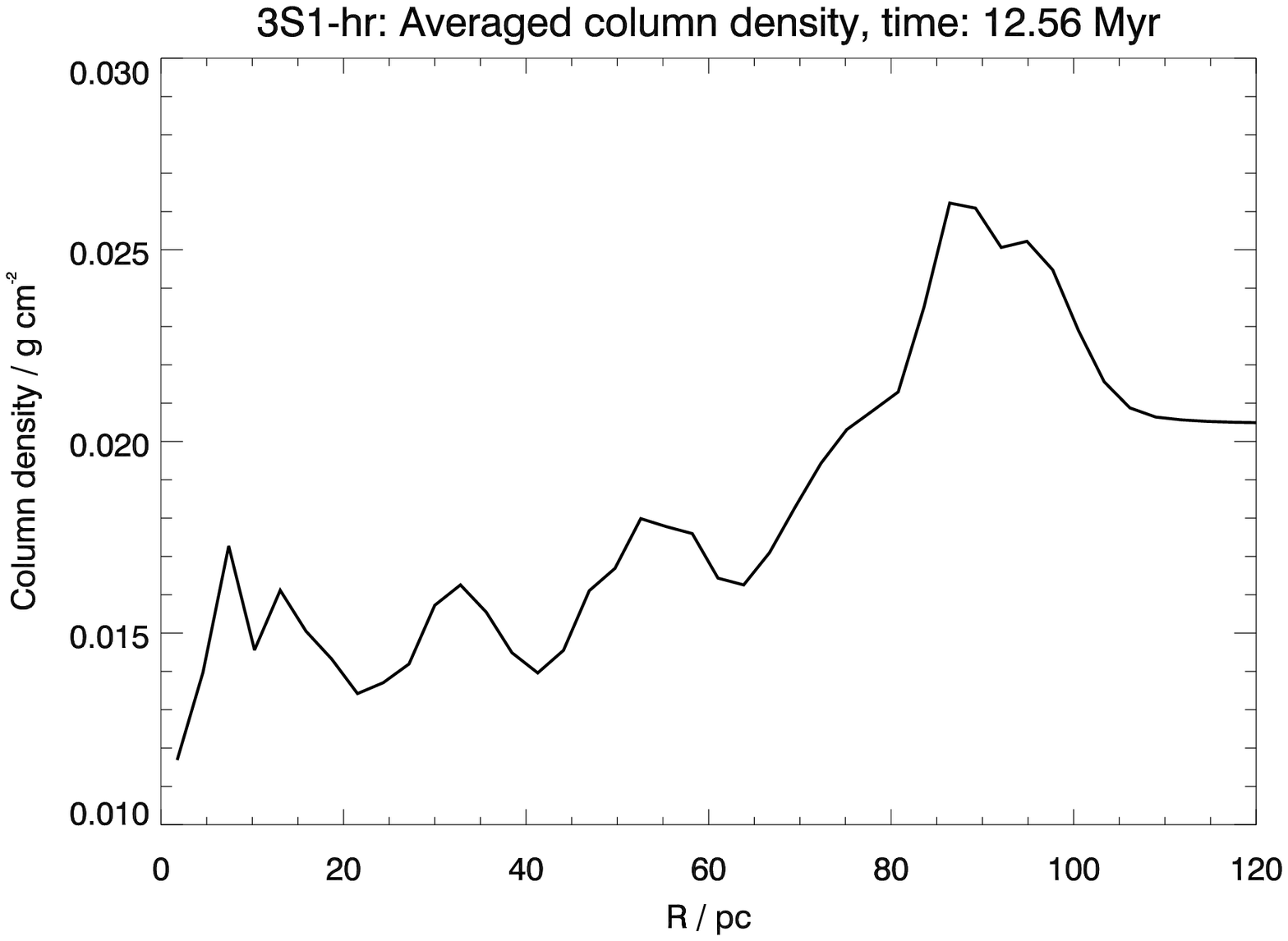}
  \includegraphics[width=0.45\textwidth]{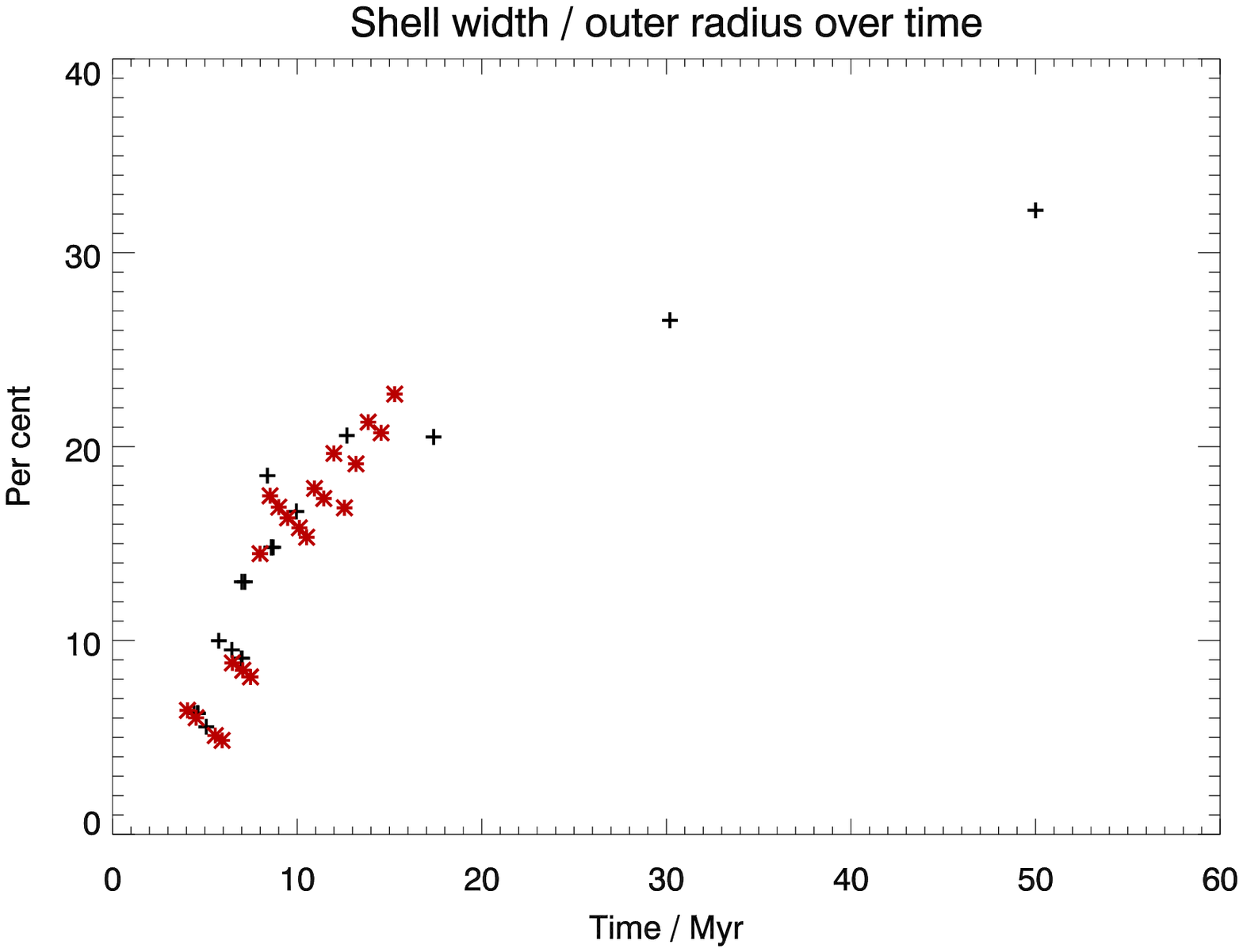}
   \includegraphics[width=0.45\textwidth]{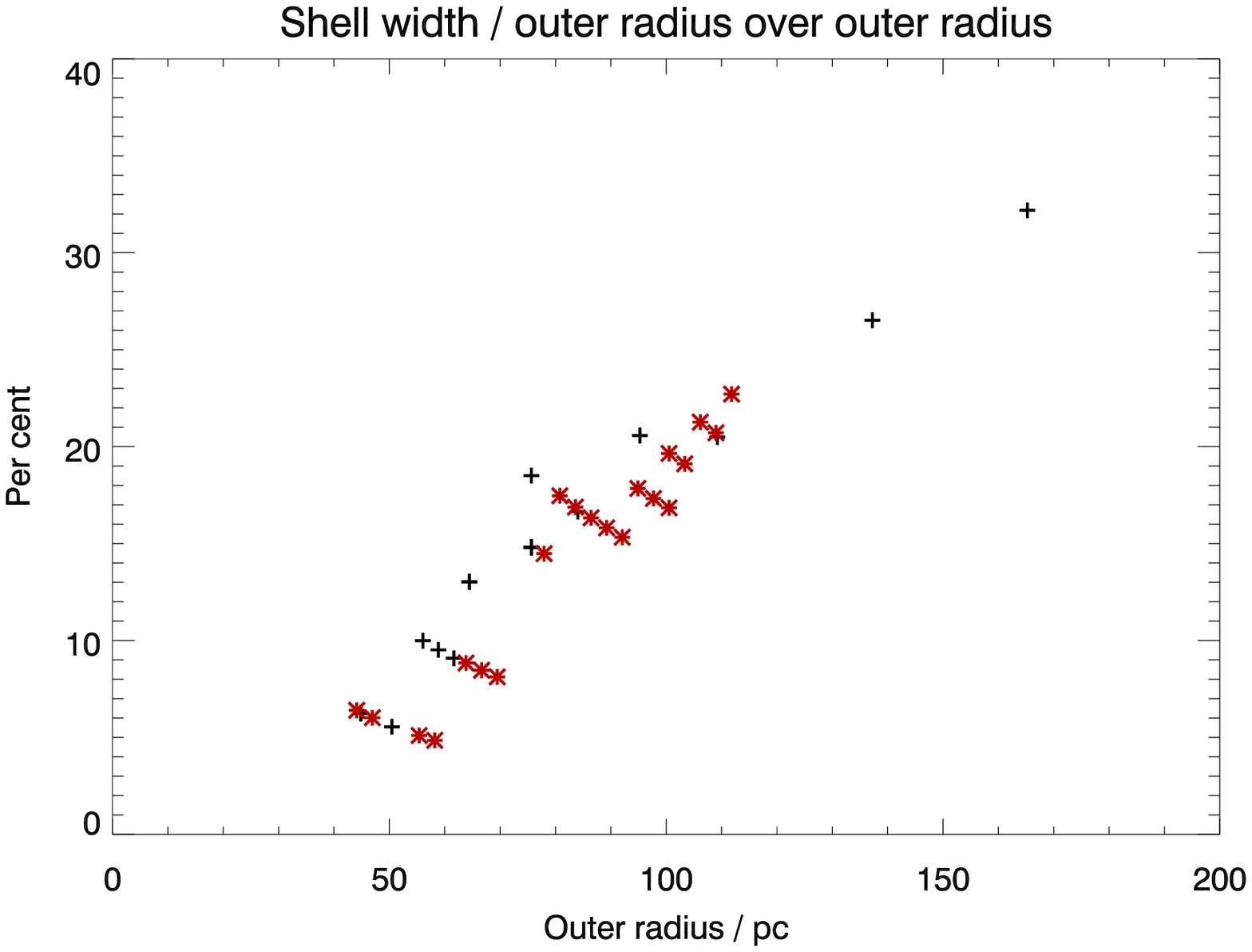}
  \caption{Shell width for column density maps. Top: Angle-averaged
  column density over radius for run 3S1-hr at 12.56~Myr. From such
  plots, the shell width has been determined as the radial range where
the column density is at least five per cent greater than at large radii
(undisturbed gas). The shell width determined in this way is shown in
the middle plot as a function of time, and in the bottom plot as a
function of outer radius. Black pluses are for run 3S1, red stars for
run 3S1-hr. The average shell width does not depend significantly on resolution.}
  \label{f:sw}%
\end{figure}

We find that our simulated shells are widened due to the Vishniac
instability. For the determination of the shell width, we average the
column density maps over the angle, and identify the shell as radial interval where
the column density is at least five per cent higher than in the undisturbed
medium. The shell width is shown in Figure ~\ref{f:sw}
as a function of time and radius, respectively, for runs 3S1 and
3S1-hr. For this analysis, we only use late snapshots, where the
superbubbles are well established.

The shell width is typically in the tens of per cent regime and
increases with time. The result does not depend on the resolution.
    
%
%
%
%
%
%
\begin{figure}
  \centering
  \includegraphics[width=0.45\textwidth]{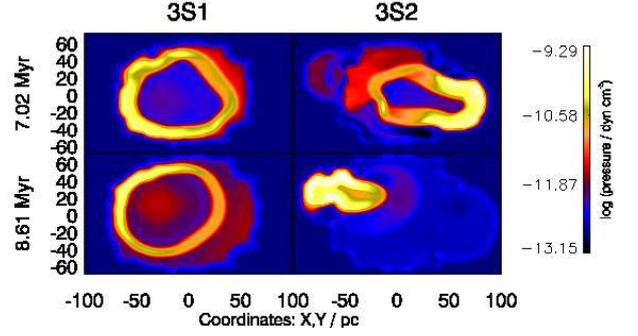}
  \caption{Pressure maps of run 3S1 (left) and 3S2 (right), shortly
    after the second (top) and third (bottom) supernova.  }
  \label{f:pmaps}%
\end{figure}
%
%


\section{Discussion}\label{s:disc}
We have investigated the environmental impact of a group of three
massive stars via 3D~hydrodynamic simulation. Herein, several
assumptions and simplifications were necessarily introduced:

We have adopted a uniform background density of
10~$m_\mathrm{p}$~\ccmi. On scales of ten pc and smaller, the density
will in reality be at least a factor of ten higher
\citep[e.g.][]{Kainulea11}. On scales of 100~pc, the density should
become equal to or even smaller than about 1~$m_\mathrm{p}$~\ccmi~
\citep[e.g.][]{dAB05}. Hence, our choice should be realistic for the
tens of pc scales we simulate \citep[compare
also][]{FHY03,FHY06,vanMea12}.  The real ISM has a rich spatial
structure, whereas we use a homogeneous distribution. This is a
significant difference. For a porous ISM, the injected wind/SN energy
could escape through low density regions making the bubbles
smaller (Fierlinger et al. 2012b\rev{, in prep.}). 
For such a situation, one should also expect pronounced
bubble asymmetries. Indeed, such asymmetries are found in observations
\citep[e.g.][]{Churchea06}. Yet, in order to be able to compare the
effect of different spatial configurations of the stars, and not to be
dominated by local environmental effects it is necessary to use a
homogeneous background density.


We have found that with the
standard ISM thermodynamics, the peak shell density does not converge
with finer resolution. It is not immediately obvious that this should
be so, as the
photo-electric heating we take into account could in principle have
produced high enough pressure to limit the shell compression. Yet,
with our highest resolution of 0.5~pc, this has not been the case.
The shell density results from several effects:
At a given pressure level, there is a density and a temperature that
correspond to thermodynamic equilibrium. When the pressure in the
bubble increases, e.g. because a supernova has happened, the shell can
however not adjust immediately to the new equilibrium pressure level, because the
gas has to be swept together in a finite time. The inverse
happens at late times (as demonstrated in Figure~\ref{f:3s1-hdzoom})
when the bubble pressure strongly decreases, but the clumps in the
shell cannot expand fast enough to remain in pressure equilibrium.
The compression is of course also limited by the resolution. The
non-convergence therefore means that the bubble pressure is high
enough for a sufficiently long time so that compression of the shell,
or some clumps therein, to even higher densities
may occur if one would repeat the simulation with an even higher
resolution. For gas on the thermodynamic equilibrium curve in the relevant density
regime, higher densities correspond to lower temperatures.
Compared to observations
\citep[e.g.][]{Preibea12}, the ISM in star forming regions rarely
reaches temperatures below about 20~K, and 20~K to 100~K are typical for the
dense phase. Similar temperatures are also found in our simulated
shells. Other effects like magnetic fields, self-gravity or feedback by the new stars,
which in reality might form in our dense clouds, may affect
the cloud compression, but are not included in our simulation.
Thus, even if the compression would increase still further if one
would carry out the simulations at yet higher resolution, this would
not necessarily be more realistic, as the high density clumps may be
regarded as physical systems of their own with some of the physics
necessary to describe them properly not being present in our simulations.

The absolute value of the energy deposition is also resolution
dependent. It increases by about a factor of 1.2 if we double the
resolution. The
reason for this is likely related to the Vishniac instability:
\citet{VR89} estimate the wavelength at which the growth rate is
largest as:
\dm{\lambda_\mathrm{VI,max} \approx 0.3\,\mathrm{pc}\;
 \left(\frac{\Sigma_0}{10^{-3}\,\mathrm{g\,cm}^{-2}}\right)
   \left(\frac{10^{-12}\,\mathrm{dyn\,cm}^{-2}}{P_\mathrm{i}}\right)
  \left( \frac{10^{-9}\,\mathrm{cm\,s}^{-2}}{a}\right)\, ,}
where we have plugged in typical values for the column density
$\Sigma_0$, the internal pressure $P_\mathrm{i}$, and the shell
deceleration $a$. This is comparable to our best
resolution. Therefore, finer resolution should still trigger strongly
unstable Vishniac modes, which seem to have an effect on the
result. The minimum unstable wavelength is predicted to be 
$0.5\lambda_\mathrm{VI,max}$. Unfortunately, for the present study we
did not have the computational resources to probe these scales, but
this should become possible in future. In contrast to the Vishniac
instability, the Rayleigh-Taylor instability continues to grow faster
for smaller wavelengths. Thermal conduction would be expected to be
important at even smaller scales of about 0.01~pc$/n$, where $n$ is the
number density in the shell \citep{MC77}. Thus, our absolute
efficiency numbers are lower limits. 

At the level of this accuracy, 3D effects might be
important, because the shell instabilities should be 3D in nature. Up
to the first supernova, our simulations are dominated by the wind of
the 60~\ms star, and may thus be compared to the 2D results of
\citet{FHY03}.  We find an energy response of at least 10~per cent,
which compares to 9~per cent in the simulation of \citet{FHY03}, which
is very similar. It might point to some effect in the direction
that more energy is retained in the ISM in 3D simulations, but could
also be related to numerical details or the slightly higher density they use.

The
general shape of the curves is however well converged (compare
Figure~\ref{f:3sw1-eres}). As a further check, we have also
resimulated run 3S0 at the resolution of 3S1-hr. The energy deposition
ratio between the two high resolution simulations is very similar to
the one 
\rev{at low resolution.}
We therefore believe that the relative trends of the
energy deposition we report here are reliable.

We find that the Vishniac instability dominates the shell
evolution. We show that the instability in our simulations 
is connected to the shells'
overdensity and to non-radial motions in the shells, in agreement with
the predictions of \citet{VR89} and \citet{MLN93}.  Limiting the
shell's overdensity by e.g. magnetic fields would therefore directly
affect the Vishniac instability. 

From the column density plots
(Figures~\ref{f:3sw1-overview}, \ref{f:3sw1-overviewb} and
\ref{f:3s1_cd_rescomp}), it is obvious that the observational
appearance of the shell is dominated by the Vishniac instability: If
the shells were smooth, and the maximum density would increase with
resolution as seen in our simulations (Figure~\ref{f:3sw1_maxd-t}),
one would expect that the shell gets thinner with finer resolution, as
the smaller cells allow higher compression. Yet,
we find a radially averaged shell width of 
tens of~per~cent of the outer radius independent of resolution 
(Figure~\ref{f:sw}). In the low resolution simulation, much of
the width is due to the large scale distortion influenced by the grid
directions. \rev{F}or the high resolution simulation the width is due to
small wavelength modes.

In their survey of 322~interstellar bubbles,
\citet{Churchea06} find typical shell widths of 20-40~per~cent of the
outer radius. Thus, it seems unlikely that the development of the
Vishniac instability is frequently impeded by anything, e.g. limited
compression due to magnetic fields, as this would again make the
shells thin. In other words, in order to study the effects of magnetic fields one probably needs much higher numerical resolution than adopted in our models.

The column density
should give a rough indication on observed morphologies. From the
corresponding plots, we find that the Vishniac instability should also
lead to observable filamentary structure inside the bubbles. This
seems to be the case for some shells associated with supernova
remnants 
\citep[e.g. Crab,][\revtwo{compare also the discussion in
\citet{vMK12}}]{He08}, 
which confirms the above
analysis. More detailed comparison would of course be interesting.

\begin{figure}
  \centering
  \includegraphics[width=0.49\textwidth]{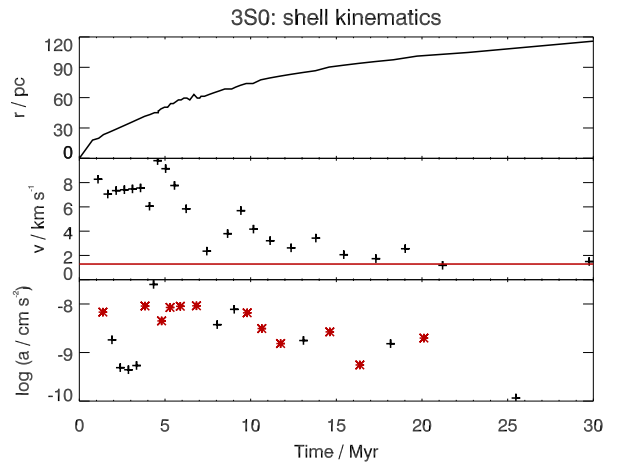}
  \caption{Shell kinematics (top: radius, middle: velocity, bottom:
    acceleration), as functions of time for run 3S0. The velocity
    points are averaged over time intervals of varying length, which
    correspond to shell radii differences of at least 2 cells. The
    shell velocity converges towards the ambient sound speed (red
    line). Each supernova leads to a significant acceleration of the
    shell (black crosses, bottom plot), followed by a comparably
    strong deceleration (red stars).}
  \label{f:3S0-rva}%
\end{figure}

We find that the best way to inject energy into the ISM, i.e. to
achieve a high energy response is a continuous, steady energy
injection. Supernovae dissipate their energy within about 1~Myr.  We
show the kinematics for run 3S0 (all stars at same position) in
Figure~\ref{f:3S0-rva}.  After each supernova, the shell accelerates
significantly. This means more kinetic energy in the shell. Yet the
increased expansion leads to fast adiabatic pressure loss of the shell
interior. The increased kinetic energy is quickly dissipated at the
leading radiative bow shock, as long as it is strongly supersonic. 
In contrast, the energy fraction deposited in the ISM in the wind phase
remains roughly constant at ten per cent.
Thus, retaining the injected energy in an interstellar bubble requires
continuous energy injection.

The energy tracks of merging bubbles are entirely dominated by these
shell kinematics effects. For example, in run 3S1, the merging process
has clearly set in at 2~Myr (compare the high resolution version,
Figure~\ref{f:3sw1-overview}) and continues for a few Myr
thereafter. Yet, the energy track for this time interval is
indistinguishable from run 3S2 (different positions of the stars) and
even from 3S0 (no shell merging because drivers are at same location)
and the sum of S25, S32 and S60 (no shell merging because the stars
are sufficiently far away, realised by having them in different
simulations).


\ch{Exploding a supernova in a
superbubble and not in its own wind bubble leads to weaker radiative
losses:}  
Each supernova shock heats first the bubble interior. It then makes a
difference how large the respective bubble is in communicating the
thermal energy to the shell: For larger bubbles, the heat energy is
distributed over a greater volume. Thus the overpressure is
smaller. The force on the shell is correspondingly smaller. Hence,
shell acceleration and adiabatic losses of the bubble interior happen
on a longer timescale. This is the reason for the longer energy decay
timescale for each subsequent supernova.  Consequently, after a
supernova, the energy decays fastest if the bubbles remain isolated,
as each star has a small bubble of its own. 

Off-centre explosions are another significant effect for the energy
tracks: The first supernova always explodes roughly in the middle of
the superbubble. This must of course be so at least for coeval stars,
since its parent star also has the highest energy output and is the
dominant driver of the superbubble before it explodes. The energy
tracks of the simulations with different \revtwo{spatial}
configurations \revtwo{of the stars} \ch{show
  little difference} up to the point when the second star
explodes. This happens necessarily significantly off-centre. The
explosion accelerates first and most efficiently the parts of the
super-shell which are most nearby (compare the pressure maps in
Figure~\ref{f:pmaps}).  Yet, if the bubbles are fully merged at the
time of the explosion (3S1) the effect is only at the per cent
level. This is due to the high sound speed within the bubble, which
communicates pressure differences quickly. We notice a considerable
effect on the energy track for run 3S2, where the individual bubbles
are still well identifiable at the time of the final supernova.

Thus, especially where the shells are not yet fully merged at the time
of explosion, the off-centre location leads to a certain extent to a
behaviour closer to the isolated bubble case.  Therefore, the energy
tracks (Figure~\ref{f:etracks-dist}) of runs 3S1 and 3S2 essentially
do not leave the range spanned by the isolated bubbles case
(S25+S32+S60) and the cospatial parent star case (3S0).

\ch{Another finding which might seem curious is that all the energy
  tracks in Figure~\ref{f:etracks-dist} converge at late times. Long
  after the energy injection has ceased, the energy of the affected
  gas is dominated by the kinetic energy of the shell. Because the
  swept up mass is dominated by the action of the $60~M_\odot$ star and the final
  shell velocity is always similar to the sound speed of the ambient
  medium, the overall energy increase is very similar in all
  simulations.}
%
%
%
%

Population synthesis of stellar groups/subgroups combined with energy
injection data from stellar evolutionary models \citep{Vossea09} show
that the wind energy dominates within the first few Myr after the
star formation event. Later, the energy input is dominated by
supernovae. Observed subgroups have an age difference of order a few
Myr \citep{Vossea10,Vossea12}. Thus, it appears possible that the
energy response (compare equation~(\ref{eq:eres})) 
is kept \ch{high} for $\gtrsim 10$~Myr by the wind
contributions of different subgroups coming in at slightly different
times.  Observations find energy responses of about ten per cent or
higher \citep[e.g.][]{OG04,Vossea12}. This agrees very well with the
results in the wind phase of our highest resolution run and might
suggest that additional effects, which are not taken into account in
our simulation and which we believe should only increase the energy
response, may not dominate. 

A similar energy response has also been
inferred observationally for galactic winds \citep[e.g.][]{VCB05}, though only
the supernova energy has been taken into account for the
calculation. Galactic winds are thought to arise as a final merging
stage from central superbubbles in star-forming galaxies. 
If one wants to keep the energy response high in order to match
  the constraints from the galactic wind observations, the individual
  bubbles should \rev{be closely spaced and} 
  merge early in order to have as constant an energy
  input rate as possible.
\rev{This is of course the case for wind galaxies, such as
  \object{M82}, with their star clusters and even super-star clusters
  \citep[e.g.][]{FSea03, Westmoquea09}.
The same effect that we observe for individual stars, namely that their energy
deposition is higher, if they are closer together, should also apply
to clusters of stars: If two clusters are closer together, they should
deposit more energy into the ISM as if they were further apart.}

\section{Conclusions}
We have simulated isolated interstellar bubbles and emerging
superbubbles which form from adjacent interstellar bubbles with
stellar distances of order tens of pc. Thus, our simulations apply,
within the limitations outlined in Section~\ref{s:disc} above, well
to hierarchically clustered star formation complexes like the Orion
\citep{Vossea10}, Scorpius-Centaurus \citep{Diehlea10} or Carina
\citep{Vossea12} regions.

We find in our simulations that \ch{up to about the second supernova} 
the total energy of superbubbles is
\ch{not strongly dependent on} the spatial configuration of the 
\ch{group of} parent stars,
including zero and infinite distance. 
Off-centre energy injection reduces the ISM energy response
significantly only, if the individual bubbles are not yet fully
merged.  Thus, from before the second supernova onwards the energy
response is higher for more closely packed configurations. We find on
average about a factor of two difference in energy response between
the isolated stars-case and the cospatial stellar configuration.


Supernovae increase the ISM energy only for very small timescales of
about 1~Myr\ch{, increasing with the size of the superbubble at the
  time of the explosion}. After that time, the retained energy is {\em smaller}
than immediately before the supernova (Figure~\ref{f:3S1-hr-e}). The
energy response drops by a factor of two shortly after the supernova
compared to the main sequence wind phase.  Our simulations are quite
realistic regarding the time intervals in between subsequent supernova
events \citep[compare][]{Vossea09}. Thus, we conclude that for
realistic star clusters energy is build up in the wind
phases. Supernovae lead to large short term energy variations, but
only keep up the bubble energy in the long run, at a roughly constant level. 

We also
find that supernovae that explode inside larger bubbles have a longer
energy decay time. The 60~\ms star has produced a bubble of $\gtrsim
80$~pc diameter at the time it explodes. Thus in order to obtain a
physically sound feedback model, which is currently lacking in studies
of disk galaxies \citep{Scanea12}, it seems essential to
account for the wind phase. Further, since the energy deposition does
essentially not depend on the spatial configuration of the stars, up
to stellar distances of about 30~pc in our simulations, it seems
reasonable to use stellar clusters as fundamental feedback units, not
individual stars, or in other words superbubbles rather than
individual bubbles of individual stars, at least for a clustered star
formation mode, which should according to our simulations be more
efficient for feedback purposes.

We have verified by comparison to theoretical work that the appearance
of our wind shells is dominated by the Vishniac instability, 
\revtwo{which is now for the first time prominently seen in 3D
simulations \citep[this article and][]{vMK12}.} High
resolution is essential to obtain the necessary shell overdensities
which are crucial for the development of the instability. This effect
widens the shell significantly in column density plots, which we
suggest may explain the large observed shell widths of 20~per cent of
the outer radii and more. It also produces filamentary structure in
the shell which is also well visible in our column density plots.
\ch{We conclude that} filamentary structure inside
interstellar bubbles may be related to the Vishniac instability.

\begin{acknowledgements}
  This research was supported by the cluster of excellence ``Origin
  and Structure of the Universe'' (www.universe-cluster.de). We thank
  the anonymous referee for a very useful report, and Mordecai-Mark
  Mac Low for very helpful comments.
\end{acknowledgements}

\bibliographystyle{aa}
\bibliography{/Users/mkrause/texinput/references}

\end{document}

%% file: def.tex
\bibpunct{(}{)}{;}{a}{}{,}
\newcommand{\ignore}[1]{}

\newcommand{\eql}[2]{\begin{equation} \label{#1} #2 \end{equation}}

\newcommand{\ccmi}{$\mathrm{cm}^{-3}$}

\newcommand{\nv}{{\sc Nirvana~3.5~}}
\newcommand{\nve}{{\sc Nirvana~3.5}}

\newcommand{\ms}{$M_\odot$ }

\newcommand{\dm}[1]{\begin{displaymath} #1 \end{displaymath}}

\newcommand{\sqrbrk}[1]{$\left[\right.$#1$\left.\right]$}